\documentclass[11pt]{article}
\input{psfig.sty}
\newcommand{\dshv}{
\begin{picture}(52,10)
\put(15,3){\line(-5,2){10}}
\put(15,3){\line(-5,-2){10}}
\multiput(15,3)(22,0){2}{\circle*{2}}
\multiput(15,3)(6,0){4}{\line(1,0){4}}
\put(37,3){\line(5,2){10}}
\put(37,3){\line(5,-2){10}}
\end{picture}  }
\newcommand{\zigv}{
\begin{picture}(50,10)
\put(15,3){\line(-5,2){10}}
\put(15,3){\line(-5,-2){10}}
\multiput(15,3)(20.15,0){2}{\circle*{2}}
\multiput(15,3)(0.42,0.42){4}{\circle*{1}}
\multiput(16.26,4.25)(0.5,-0.5){6}{\circle*{1}}
\multiput(18.76,1.75)(0.5,0.5){6}{\circle*{1}}
\multiput(21.26,4.25)(0.5,-0.5){6}{\circle*{1}}
\multiput(23.76,1.75)(0.5,0.5){6}{\circle*{1}}
\multiput(26.26,4.25)(0.5,-0.5){6}{\circle*{1}}
\multiput(28.76,1.75)(0.5,0.5){6}{\circle*{1}}
\multiput(31.26,4.25)(0.5,-0.5){6}{\circle*{1}}
\multiput(33.76,1.75)(0.42,0.42){4}{\circle*{1}}
\put(35.15,3){\line(5,2){10}}
\put(35.15,3){\line(5,-2){10}}
\end{picture}  }
\newcommand{\dotv}{
\begin{picture}(33,10)
\multiput(6,3)(4,0){3}{\circle*{2}}
\put(18,3){\circle*{3}}
\put(18,3){\line(5,2){10}}
\put(18,3){\line(5,-2){10}}
\end{picture}  }
\newcommand{\ddotv}{
\begin{picture}(54,10)
\put(15,3){\line(-5,2){10}}
\put(15,3){\line(-5,-2){10}}
\multiput(15,3)(24,0){2}{\circle*{3}}
\multiput(19,3)(4,0){5}{\circle*{2}}
\put(39,3){\line(5,2){10}}
\put(39,3){\line(5,-2){10}}
\end{picture}  }
\newcommand{\Gausd}{
\begin{picture}(55,10)
\thicklines
\put(6,-1){\line(1,0){40}}
\put(6,1){{\bf k}}
\put(35,1){-{\bf k}}
\end{picture}  }
\newcommand{\dashd}{
\begin{picture}(66,15)
\thicklines
\put(6,-1){\line(1,0){54}}
\put(6,1){{\bf k}}
\put(49,1){-{\bf k}}
\multiput(18,-1)(28,0){2}{\circle*{3}}
\multiput(18,-1)(0.05,0.5){9}{\circle*{1}}
\multiput(46,-1)(-0.05,0.5){9}{\circle*{1}}
\multiput(19,5)(0.3,0.475){9}{\circle*{1}}
\multiput(45,5)(-0.3,0.475){9}{\circle*{1}}
\multiput(23.4,10.6)(0.45,0.25){9}{\circle*{1}}
\multiput(40.6,10.6)(-0.45,0.25){9}{\circle*{1}}
\multiput(30,13.2)(0.25,0){17}{\circle*{1}}
\end{picture}  }
\newcommand{\zigd}{
\begin{picture}(66,15)
\thicklines
\put(6,-1){\line(1,0){54}}
\put(6,1){{\bf k}}
\put(49,1){-{\bf k}}
\multiput(18,-1)(26.6,0){2}{\circle*{3}}
\multiput(18,-1)(-0.1,0.4){11}{\circle*{1}}
\multiput(44.6,-1)(0.1,0.4){11}{\circle*{1}}
\multiput(17,3)(0.4,-0.05){11}{\circle*{1}}
\multiput(45.6,3)(-0.4,-0.05){11}{\circle*{1}}
\multiput(21,2.5)(-0.1,0.4){11}{\circle*{1}}
\multiput(41.6,2.5)(0.1,0.4){11}{\circle*{1}}
\multiput(20,6.5)(0.4,-0.1){11}{\circle*{1}}
\multiput(42.6,6.5)(-0.4,-0.1){11}{\circle*{1}}
\multiput(24,5.5)(0,0.4){11}{\circle*{1}}
\multiput(38.6,5.5)(0,0.4){11}{\circle*{1}}
\multiput(24,9.5)(0.35,-0.22){11}{\circle*{1}}
\multiput(38.6,9.5)(-0.35,-0.22){11}{\circle*{1}}
\multiput(27.5,7.3)(0.15,0.37){11}{\circle*{1}}
\multiput(35.1,7.3)(-0.15,0.37){11}{\circle*{1}}
\multiput(29,11)(0.23,-0.32){11}{\circle*{1}}
\multiput(33.6,11)(-0.23,-0.32){11}{\circle*{1}}
\end{picture}  }
\newcommand{\dotd}{
\begin{picture}(66,15)
\thicklines
\put(6,-1){\line(1,0){54}}
\put(6,1){{\bf k}}
\put(49,1){-{\bf k}}
\multiput(18,-1)(28,0){2}{\circle*{3}}
\multiput(18.5,3)(27,0){2}{\circle*{2}}
\multiput(20.6,7)(22.8,0){2}{\circle*{2}}
\multiput(24,10)(16,0){2}{\circle*{2}}
\multiput(28,12)(8,0){2}{\circle*{2}}
\put(32,12.6){\circle*{2}}
\end{picture}  }
\newcommand{\bldot}{
\begin{picture}(32,12)
\put(9,3.5){\circle{11}}
\multiput(8.8,-1.6)(0.5,0.5){12}{\circle*{1}}
\multiput(9.2,8.6)(-0.5,-0.5){12}{\circle*{1}}
\multiput(6.2,-1)(0.5,0.5){15}{\circle*{1}}
\multiput(11.8,8)(-0.5,-0.5){15}{\circle*{1}}
\multiput(14.5,3.5)(3.5,0){5}{\circle*{2}}
\end{picture}  }
\newcommand{\blddot}{
\begin{picture}(28.5,12)
\put(9,3.5){\circle{11}}
\multiput(8.8,-1.6)(0.5,0.5){12}{\circle*{1}}
\multiput(9.2,8.6)(-0.5,-0.5){12}{\circle*{1}}
\multiput(6.2,-1)(0.5,0.5){15}{\circle*{1}}
\multiput(11.8,8)(-0.5,-0.5){15}{\circle*{1}}
\multiput(14.3,6)(3.5,0.75){4}{\circle*{2}}
\multiput(14.3,1)(3.5,-0.75){4}{\circle*{2}}
\end{picture}  }
\newcommand{\bltdot}{
\begin{picture}(28.5,15)
\put(9,3.5){\circle{11}}
\multiput(8.8,-1.6)(0.5,0.5){12}{\circle*{1}}
\multiput(9.2,8.6)(-0.5,-0.5){12}{\circle*{1}}
\multiput(6.2,-1)(0.5,0.5){15}{\circle*{1}}
\multiput(11.8,8)(-0.5,-0.5){15}{\circle*{1}}
\multiput(14,7)(3.5,0.8){4}{\circle*{2}}
\multiput(14,0)(3.5,-0.8){4}{\circle*{2}}
\multiput(14.5,3.5)(3.5,0){4}{\circle*{2}}
\end{picture}  }
\newcommand{\blzig}{
\begin{picture}(32.5,12)
\put(9,3.5){\circle{11}}
\multiput(8.8,-1.6)(0.5,0.5){12}{\circle*{1}}
\multiput(9.2,8.6)(-0.5,-0.5){12}{\circle*{1}}
\multiput(6.2,-1)(0.5,0.5){15}{\circle*{1}}
\multiput(11.8,8)(-0.5,-0.5){15}{\circle*{1}}
\multiput(14.5,3.5)(0.5,0.5){3}{\circle*{1}}
\multiput(15.5,4.5)(0.5,-0.5){5}{\circle*{1}}
\multiput(17.5,2.5)(0.5,0.5){5}{\circle*{1}}
\multiput(19.5,4.5)(0.5,-0.5){5}{\circle*{1}}
\multiput(21.5,2.5)(0.5,0.5){5}{\circle*{1}}
\multiput(23.5,4.5)(0.5,-0.5){5}{\circle*{1}}
\multiput(25.5,2.5)(0.5,0.5){5}{\circle*{1}}
\multiput(27.5,4.5)(0.5,-0.5){5}{\circle*{1}}
\end{picture}  }
\newcommand{\bldzig}{
\begin{picture}(30,12)
\put(9,3.5){\circle{11}}
\multiput(8.8,-1.6)(0.5,0.5){12}{\circle*{1}}
\multiput(9.2,8.6)(-0.5,-0.5){12}{\circle*{1}}
\multiput(6.2,-1)(0.5,0.5){15}{\circle*{1}}
\multiput(11.8,8)(-0.5,-0.5){15}{\circle*{1}}
\multiput(14.2,5.5)(0.4,0.55){5}{\circle*{1}}
\multiput(15.8,7.7)(0.55,-0.4){5}{\circle*{1}}
\multiput(18,6.1)(0.4,0.55){5}{\circle*{1}}
\multiput(19.6,8.3)(0.55,-0.4){5}{\circle*{1}}
\multiput(21.8,6.7)(0.4,0.55){5}{\circle*{1}}
\multiput(23.4,8.9)(0.55,-0.4){5}{\circle*{1}}
\multiput(25.6,7.3)(0.32,0.44){4}{\circle*{1}}
\multiput(14.2,1.5)(0.4,-0.55){5}{\circle*{1}}
\multiput(15.8,-0.7)(0.55,0.4){5}{\circle*{1}}
\multiput(18,0.9)(0.4,-0.55){5}{\circle*{1}}
\multiput(19.6,-1.3)(0.55,0.4){5}{\circle*{1}}
\multiput(21.8,0.3)(0.4,-0.55){5}{\circle*{1}}
\multiput(23.4,-1.9)(0.55,0.4){5}{\circle*{1}}
\multiput(25.6,-0.3)(0.32,-0.44){4}{\circle*{1}}
\end{picture}  }
\newcommand{\bltzig}{
\begin{picture}(30,15)
\put(9,3.5){\circle{11}}
\multiput(8.8,-1.6)(0.5,0.5){12}{\circle*{1}}
\multiput(9.2,8.6)(-0.5,-0.5){12}{\circle*{1}}
\multiput(6.2,-1)(0.5,0.5){15}{\circle*{1}}
\multiput(11.8,8)(-0.5,-0.5){15}{\circle*{1}}
\multiput(13.7,7)(0.4,0.55){5}{\circle*{1}}
\multiput(15.3,9.2)(0.55,-0.4){5}{\circle*{1}}
\multiput(17.5,7.6)(0.4,0.55){5}{\circle*{1}}
\multiput(19.1,9.8)(0.55,-0.4){5}{\circle*{1}}
\multiput(21.3,8.2)(0.4,0.55){5}{\circle*{1}}
\multiput(22.9,10.4)(0.55,-0.4){5}{\circle*{1}}
\multiput(25.1,8.8)(0.32,0.44){4}{\circle*{1}}
\multiput(13.7,0)(0.4,-0.55){5}{\circle*{1}}
\multiput(15.3,-2.2)(0.55,0.4){5}{\circle*{1}}
\multiput(17.5,-0.6)(0.4,-0.55){5}{\circle*{1}}
\multiput(19.1,-2.8)(0.55,0.4){5}{\circle*{1}}
\multiput(21.3,-1.2)(0.4,-0.55){5}{\circle*{1}}
\multiput(22.9,-3.4)(0.55,0.4){5}{\circle*{1}}
\multiput(25.1,-1.8)(0.32,-0.44){4}{\circle*{1}}
\multiput(14.5,3.5)(0.5,0.5){3}{\circle*{1}}
\multiput(15.5,4.5)(0.5,-0.5){5}{\circle*{1}}
\multiput(17.5,2.5)(0.5,0.5){5}{\circle*{1}}
\multiput(19.5,4.5)(0.5,-0.5){5}{\circle*{1}}
\multiput(21.5,2.5)(0.5,0.5){5}{\circle*{1}}
\multiput(23.5,4.5)(0.5,-0.5){5}{\circle*{1}}
\multiput(25.5,2.5)(0.5,0.5){3}{\circle*{1}}
\end{picture}  }
\newcommand{\selfe}{
\begin{picture}(48,10)
\thicklines
\put(24,3.5){\circle{11}}
\multiput(23.8,-1.6)(0.5,0.5){12}{\circle*{1}}
\multiput(24.2,8.6)(-0.5,-0.5){12}{\circle*{1}}
\multiput(21.2,-1)(0.5,0.5){15}{\circle*{1}}
\multiput(26.8,8)(-0.5,-0.5){15}{\circle*{1}}
\put(18.5,3.5){\line(-1,0){15}}
\put(29.5,3.5){\line(1,0){15}}
\end{picture}  }
\newcommand{\skd}{
\begin{picture}(45,10)
\multiput(9,3.5)(27,0){2}{\circle{11}}
\multiput(14.5,3.5)(16,0){2}{\circle*{3}}
\multiput(14.5,3.5)(6,0){3}{\line(1,0){4}}
\end{picture}  }
\newcommand{\dg}{
\begin{picture}(55,14)
\put(27.5,3.5){\circle{16}}
\multiput(19.5,3.5)(16,0){2}{\circle*{3}}
\multiput(19.5,3.5)(-6,0){3}{\line(-1,0){4}}
\multiput(35.5,3.5)(6,0){3}{\line(1,0){4}}
\put(3.5,7.5){\footnotesize {\bf q}}
\put(43,7.5){\footnotesize -{\bf q}}
\end{picture}  }
\newcommand{\ddg}{
\begin{picture}(63,17)
\put(31.5,3.5){\circle{24}}
\multiput(19.5,3.5)(24,0){2}{\circle*{3}}
\multiput(19.5,3.5)(-6,0){3}{\line(-1,0){4}}
\multiput(43.5,3.5)(6,0){3}{\line(1,0){4}}
\put(3.5,7.5){\footnotesize {\bf q}}
\put(50.5,7.5){\footnotesize -{\bf q}}
\multiput(31.5,15.5)(0,-24){2}{\circle*{3}}
\multiput(31.7,12.3)(0,-7.2){3}{\oval(4,4)[tl]}
\multiput(31.7,12.7)(0,-7.2){3}{\oval(4,4)[bl]}
\multiput(31.3,8.7)(0,-7.2){3}{\oval(4,4)[tr]}
\multiput(31.3,9.1)(0,-7.2){3}{\oval(4,4)[br]}
\end{picture}  }
\newcommand{\dddg}{
\begin{picture}(79,17)
\put(39.5,3.5){\oval(40,24)}
\multiput(19.5,3.5)(40,0){2}{\circle*{3}}
\multiput(19.5,3.5)(-6,0){3}{\line(-1,0){4}}
\multiput(59.5,3.5)(6,0){3}{\line(1,0){4}}
\put(3.5,7.5){\footnotesize {\bf q}}
\put(66.5,7.5){\footnotesize -{\bf q}}
\multiput(31.5,15.5)(0,-24){2}{\circle*{3}}
\multiput(31.7,12.3)(0,-7.2){3}{\oval(4,4)[tl]}
\multiput(31.7,12.7)(0,-7.2){3}{\oval(4,4)[bl]}
\multiput(31.3,8.7)(0,-7.2){3}{\oval(4,4)[tr]}
\multiput(31.3,9.1)(0,-7.2){3}{\oval(4,4)[br]}
\multiput(47.5,15.5)(0,-24){2}{\circle*{3}}
\multiput(47.7,12.3)(0,-7.2){3}{\oval(4,4)[tl]}
\multiput(47.7,12.7)(0,-7.2){3}{\oval(4,4)[bl]}
\multiput(47.3,8.7)(0,-7.2){3}{\oval(4,4)[tr]}
\multiput(47.3,9.1)(0,-7.2){3}{\oval(4,4)[br]}
\end{picture}  }
\newcommand{\dddgg}{
\begin{picture}(87,17)
\put(43.5,3.5){\oval(48,24)}
\multiput(19.5,3.5)(48,0){2}{\circle*{3}}
\multiput(19.5,3.5)(-6,0){3}{\line(-1,0){4}}
\multiput(67.5,3.5)(6,0){3}{\line(1,0){4}}
\put(3.5,7.5){\footnotesize {\bf q}}
\put(74.5,7.5){\footnotesize -{\bf q}}
\multiput(31.5,15.5)(0,-24){2}{\circle*{3}}
\put(31.5,15.5){\line(0,-1){2}}
\multiput(35.5,13.9)(8,-7.2){3}{\oval(8,8)[bl]}
\multiput(35.5,5.9)(8,-7.2){3}{\oval(8,8)[tr]}
\multiput(55.5,15.5)(0,-24){2}{\circle*{3}}
\multiput(51.5,15.5)(-8,-7.2){3}{\oval(8,8)[br]}
\multiput(51.5,7.5)(-8,-7.2){3}{\oval(8,8)[tl]}
\put(31.5,-8.5){\line(0,1){2}}
\end{picture}  }
\newcommand{\dddggg}{
\begin{picture}(67,19)
\put(33.5,3.5){\circle{28}}
\multiput(19.5,3.5)(28,0){2}{\circle*{3}}
\multiput(19.5,3.5)(-6,0){3}{\line(-1,0){4}}
\multiput(47.5,3.5)(6,0){3}{\line(1,0){4}}
\put(3.5,7.5){\footnotesize {\bf q}}
\put(54.5,7.5){\footnotesize -{\bf q}}
\multiput(33.5,17.5)(0,-28){2}{\circle*{3}}
\multiput(33.7,14.1)(0,-7.2){4}{\oval(4,4)[tl]}
\multiput(33.7,14.5)(0,-7.2){4}{\oval(4,4)[bl]}
\multiput(33.3,10.5)(0,-7.2){3}{\oval(4,4)[tr]}
\multiput(33.3,10.9)(0,-7.2){3}{\oval(4,4)[br]}
\multiput(27.75,-2.5)(8,0){3}{\oval(4,4)[t]}
\multiput(23.75,-2.1)(8,0){3}{\oval(4,4)[b]}
\multiput(20.85,-2.5)(25.3,0){2}{\circle*{3}}
\end{picture}  }
\newcommand{\dddgggg}{
\begin{picture}(93.14,15)
\multiput(29.5,3.5)(34.14,0){2}{\circle{20}}
\multiput(19.5,3.5)(54.14,0){2}{\circle*{3}}
\multiput(19.5,3.5)(-6,0){3}{\line(-1,0){4}}
\multiput(73.64,3.5)(6,0){3}{\line(1,0){4}}
\put(3.5,7.5){\footnotesize {\bf q}}
\put(80.64,7.5){\footnotesize -{\bf q}}
\multiput(36.57,10.57)(20,0){2}{\circle*{3}}
\multiput(36.57,-3.57)(20,0){2}{\circle*{3}}
\multiput(38.57,10.57)(8,0){3}{\oval(4,4)[t]}
\multiput(42.57,10.57)(8,0){2}{\oval(4,4)[b]}
\multiput(38.57,-3.57)(8,0){3}{\oval(4,4)[b]}
\multiput(42.57,-3.57)(8,0){2}{\oval(4,4)[t]}
\end{picture}  }

\newcommand{\wblock}{
\begin{picture}(50,10)
\put(25,3.5){\circle{11}}
\multiput(24.8,-1.6)(0.5,0.5){12}{\circle*{1}}
\multiput(25.2,8.6)(-0.5,-0.5){12}{\circle*{1}}
\multiput(22.2,-1)(0.5,0.5){15}{\circle*{1}}
\multiput(27.8,8)(-0.5,-0.5){15}{\circle*{1}}
\multiput(17.1,3.5)(-8,0){2}{\oval(4,4)[t]}
\multiput(13.1,3.5)(-8,0){2}{\oval(4,4)[b]}
\multiput(32.9,3.5)(8,0){2}{\oval(4,4)[t]}
\multiput(36.9,3.5)(8,0){2}{\oval(4,4)[b]}
\end{picture}  }

\newcommand{\selfek}{
\begin{picture}(54,14)
\thicklines
\put(26,3.5){\circle{11}}
\multiput(25.8,-1.6)(0.5,0.5){12}{\circle*{1}}
\multiput(26.2,8.6)(-0.5,-0.5){12}{\circle*{1}}
\multiput(23.2,-1)(0.5,0.5){15}{\circle*{1}}
\multiput(28.8,8)(-0.5,-0.5){15}{\circle*{1}}
\put(20.5,3.5){\line(-1,0){17}}
\put(31.5,3.5){\line(1,0){19}}
\put(3.5,5.5){\footnotesize {\bf k}}
\put(42,5.5){\footnotesize -{\bf k}}
\multiput(12,6.5)(0.25,-0.5){13}{\circle*{1}}
\multiput(12,0.5)(0.25,0.5){13}{\circle*{1}}
\multiput(40,6.5)(-0.25,-0.5){13}{\circle*{1}}
\multiput(40,0.5)(-0.25,0.5){13}{\circle*{1}}
\end{picture} }

\newcommand{\selfch}{
\begin{picture}(117,14)
\thicklines
\put(26,3.5){\circle{11}}
\multiput(25.8,-1.6)(0.5,0.5){12}{\circle*{1}}
\multiput(26.2,8.6)(-0.5,-0.5){12}{\circle*{1}}
\multiput(23.2,-1)(0.5,0.5){15}{\circle*{1}}
\multiput(28.8,8)(-0.5,-0.5){15}{\circle*{1}}
\put(46,3.5){\circle{11}}
\multiput(45.8,-1.6)(0.5,0.5){12}{\circle*{1}}
\multiput(46.2,8.6)(-0.5,-0.5){12}{\circle*{1}}
\multiput(43.2,-1)(0.5,0.5){15}{\circle*{1}}
\multiput(48.8,8)(-0.5,-0.5){15}{\circle*{1}}
\put(89.5,3.5){\circle{11}}
\multiput(89.3,-1.6)(0.5,0.5){12}{\circle*{1}}
\multiput(89.7,8.6)(-0.5,-0.5){12}{\circle*{1}}
\multiput(86.7,-1)(0.5,0.5){15}{\circle*{1}}
\multiput(92.3,8)(-0.5,-0.5){15}{\circle*{1}}
\put(20.5,3.5){\line(-1,0){17}}
\put(31.5,3.5){\line(1,0){9}}
\put(51.5,3.5){\line(1,0){8}}
\put(63,2.5){...}
\put(76,3.5){\line(1,0){8}}
\put(3.5,5.5){\footnotesize {\bf k}}
\put(95,3.5){\line(1,0){19}}
\put(105.5,5.5){\footnotesize -{\bf k}}
\multiput(12,6.5)(0.25,-0.5){13}{\circle*{1}}
\multiput(12,0.5)(0.25,0.5){13}{\circle*{1}}
\multiput(103.5,6.5)(-0.25,-0.5){13}{\circle*{1}}
\multiput(103.5,0.5)(-0.25,0.5){13}{\circle*{1}}
\end{picture} }

\newcommand{\selftwo}{
\begin{picture}(74,14)
\thicklines
\put(26,3.5){\circle{11}}
\multiput(25.8,-1.6)(0.5,0.5){12}{\circle*{1}}
\multiput(26.2,8.6)(-0.5,-0.5){12}{\circle*{1}}
\multiput(23.2,-1)(0.5,0.5){15}{\circle*{1}}
\multiput(28.8,8)(-0.5,-0.5){15}{\circle*{1}}
\put(46,3.5){\circle{11}}
\multiput(45.8,-1.6)(0.5,0.5){12}{\circle*{1}}
\multiput(46.2,8.6)(-0.5,-0.5){12}{\circle*{1}}
\multiput(43.2,-1)(0.5,0.5){15}{\circle*{1}}
\multiput(48.8,8)(-0.5,-0.5){15}{\circle*{1}}
\put(20.5,3.5){\line(-1,0){17}}
\put(31.5,3.5){\line(1,0){9}}
\put(3.5,5.5){\footnotesize {\bf k}}
\put(51.5,3.5){\line(1,0){19}}
\put(61.5,5.5){\footnotesize -{\bf k}}
\multiput(12,6.5)(0.25,-0.5){13}{\circle*{1}}
\multiput(12,0.5)(0.25,0.5){13}{\circle*{1}}
\multiput(59.5,6.5)(-0.25,-0.5){13}{\circle*{1}}
\multiput(59.5,0.5)(-0.25,0.5){13}{\circle*{1}}
\end{picture} }
\newcommand{\sked}{
\begin{picture}(44,23)
\put(20.5,-6){\circle*{3}}
{\thicklines
\put(20.5,-6){\line(-1,0){17}}
\put(20.5,-6){\line(1,0){20}}}
\put(3.5,-4){\bf k}
\put(31,-4){-{\bf k}}
\multiput(12,-3)(0.25,-0.5){13}{\circle*{1}}
\multiput(12,-9)(0.25,0.5){13}{\circle*{1}}
\multiput(29,-3)(-0.25,-0.5){13}{\circle*{1}}
\multiput(29,-9)(-0.25,0.5){13}{\circle*{1}}
\multiput(20.5,-6)(0,6){3}{\line(0,1){4}}
\put(20.5,16){\circle{12}}
\put(20.5,10){\circle*{3}}
\end{picture} }
\newcommand{\skedd}{
\begin{picture}(72,14)
\multiput(20.5,-1)(28,0){2}{\circle*{3}}
{\thicklines
\put(20.5,-1){\line(-1,0){17}}
\put(48.5,-1){\line(1,0){20}}}
\put(3.5,1){\bf k}
\put(59,1){-{\bf k}}
\multiput(12,2)(0.25,-0.5){13}{\circle*{1}}
\multiput(12,-4)(0.25,0.5){13}{\circle*{1}}
\multiput(57,2)(-0.25,-0.5){13}{\circle*{1}}
\multiput(57,-4)(-0.25,0.5){13}{\circle*{1}}
\multiput(20.5,-1)(6,0){5}{\line(1,0){4}}
\put(34.5,-1){\oval(28,28)[t]}
\end{picture} }

\newcommand{\skbb}{
\begin{picture}(45,10)
\multiput(9,3.5)(27,0){2}{\circle{11}}
\multiput(14.5,3.5)(6,0){3}{\line(1,0){4}}
\multiput(8.8,-1.6)(0.5,0.5){12}{\circle*{1}}
\multiput(9.2,8.6)(-0.5,-0.5){12}{\circle*{1}}
\multiput(6.2,-1)(0.5,0.5){15}{\circle*{1}}
\multiput(11.8,8)(-0.5,-0.5){15}{\circle*{1}}
\multiput(35.8,-1.6)(0.5,0.5){12}{\circle*{1}}
\multiput(36.2,8.6)(-0.5,-0.5){12}{\circle*{1}}
\multiput(33.2,-1)(0.5,0.5){15}{\circle*{1}}
\multiput(38.8,8)(-0.5,-0.5){15}{\circle*{1}}
\end{picture} }
\newcommand{\skdd}{
\begin{picture}(34,16)
\put(17,3){\circle{24}}
\multiput(5,3)(24,0){2}{\circle*{3}}
\multiput(5,3)(6.5,0){4}{\line(1,0){4.5}}
\end{picture} }
\newcommand{\ssb}{
\begin{picture}(50,10)
\put(25,3.5){\circle{11}}
\multiput(24.8,-1.6)(0.5,0.5){12}{\circle*{1}}
\multiput(25.2,8.6)(-0.5,-0.5){12}{\circle*{1}}
\multiput(22.2,-1)(0.5,0.5){15}{\circle*{1}}
\multiput(27.8,8)(-0.5,-0.5){15}{\circle*{1}}
\multiput(19.5,3.5)(-6,0){3}{\line(-1,0){4}}
\multiput(30.5,3.5)(6,0){3}{\line(1,0){4}}
\end{picture} }
\newcommand{\ssbk}{
\begin{picture}(50,14)
\put(25,3.5){\circle{11}}
\multiput(24.8,-1.6)(0.5,0.5){12}{\circle*{1}}
\multiput(25.2,8.6)(-0.5,-0.5){12}{\circle*{1}}
\multiput(22.2,-1)(0.5,0.5){15}{\circle*{1}}
\multiput(27.8,8)(-0.5,-0.5){15}{\circle*{1}}
\multiput(19.5,3.5)(-6,0){3}{\line(-1,0){4}}
\multiput(30.5,3.5)(6,0){3}{\line(1,0){4}}
\put(3.5,5.5){\footnotesize {\bf k}}
\put(38,5.5){\footnotesize -{\bf k}}
\end{picture} }
\newcommand{\ssbq}{
\begin{picture}(50,14)
\put(25,3.5){\circle{11}}
\multiput(24.8,-1.6)(0.5,0.5){12}{\circle*{1}}
\multiput(25.2,8.6)(-0.5,-0.5){12}{\circle*{1}}
\multiput(22.2,-1)(0.5,0.5){15}{\circle*{1}}
\multiput(27.8,8)(-0.5,-0.5){15}{\circle*{1}}
\multiput(19.5,3.5)(-6,0){3}{\line(-1,0){4}}
\multiput(30.5,3.5)(6,0){3}{\line(1,0){4}}
\put(3.5,7.5){\footnotesize {\bf q}}
\put(38,7.5){\footnotesize -{\bf q}}
\end{picture} }
\newcommand{\ssbkn}{
\begin{picture}(50,14)
\put(25,3.5){\circle{11}}
\multiput(24.8,-1.6)(0.5,0.5){12}{\circle*{1}}
\multiput(25.2,8.6)(-0.5,-0.5){12}{\circle*{1}}
\multiput(22.2,-1)(0.5,0.5){15}{\circle*{1}}
\multiput(27.8,8)(-0.5,-0.5){15}{\circle*{1}}
\multiput(19.5,3.5)(11,0){2}{\circle*{3}}
\multiput(19.5,3.5)(-6,0){3}{\line(-1,0){4}}
\multiput(30.5,3.5)(6,0){3}{\line(1,0){4}}
\put(3.5,5.5){\footnotesize {\bf k}}
\put(38,5.5){\footnotesize -{\bf k}}
\put(15,6.5){\scriptsize 1}
\put(31,6.5){\scriptsize 2}
\end{picture} }
\newcommand{\ssch}{
\begin{picture}(124.5,14)
\put(25,3.5){\circle{11}}
\multiput(24.8,-1.6)(0.5,0.5){12}{\circle*{1}}
\multiput(25.2,8.6)(-0.5,-0.5){12}{\circle*{1}}
\multiput(22.2,-1)(0.5,0.5){15}{\circle*{1}}
\multiput(27.8,8)(-0.5,-0.5){15}{\circle*{1}}
\multiput(19.5,3.5)(-6,0){3}{\line(-1,0){4}}
\multiput(30.5,3.5)(6,0){3}{\line(1,0){4}}
\put(52,3.5){\circle{11}}
\multiput(51.8,-1.6)(0.5,0.5){12}{\circle*{1}}
\multiput(52.2,8.6)(-0.5,-0.5){12}{\circle*{1}}
\multiput(49.2,-1)(0.5,0.5){15}{\circle*{1}}
\multiput(54.8,8)(-0.5,-0.5){15}{\circle*{1}}
\multiput(57.5,3.5)(6,0){2}{\line(1,0){4}}
\put(71,2.5){...}
\multiput(84,3.5)(6,0){2}{\line(1,0){4}}
\put(99.5,3.5){\circle{11}}
\multiput(99.3,-1.6)(0.5,0.5){12}{\circle*{1}}
\multiput(99.7,8.6)(-0.5,-0.5){12}{\circle*{1}}
\multiput(96.7,-1)(0.5,0.5){15}{\circle*{1}}
\multiput(102.3,8)(-0.5,-0.5){15}{\circle*{1}}
\multiput(105,3.5)(6,0){3}{\line(1,0){4}}
\put(3.5,5.5){\footnotesize {\bf k}}
\put(112.5,5.5){\footnotesize -{\bf k}}
\end{picture} }
\newcommand{\ssbl}{
\begin{picture}(50,14)
\put(25,3.5){\circle{11}}
\multiput(19.5,3.5)(11,0){2}{\circle*{3}}
\multiput(19.5,3.5)(-6,0){3}{\line(-1,0){4}}
\multiput(30.5,3.5)(6,0){3}{\line(1,0){4}}
\put(3.5,5.5){\footnotesize {\bf k}}
\put(38,5.5){\footnotesize -{\bf k}}
\end{picture} }
\newcommand{\ssbe}{
\begin{picture}(50,14)
\put(25,3.5){\circle{11}}
\multiput(19.5,3.5)(11,0){2}{\circle*{3}}
\multiput(19.5,3.5)(-6,0){3}{\line(-1,0){4}}
\multiput(30.5,3.5)(6,0){3}{\line(1,0){4}}
\end{picture} }
\newcommand{\ssbll}{
\begin{picture}(55,14)
\put(27.5,3.5){\circle{16}}
\multiput(19.5,3.5)(16,0){2}{\circle*{3}}
\multiput(19.5,3.5)(-6,0){3}{\line(-1,0){4}}
\multiput(35.5,3.5)(6,0){3}{\line(1,0){4}}
\put(3.5,5.5){\footnotesize {\bf k}}
\put(42.5,5.5){\footnotesize -{\bf k}}
\multiput(27.5,11.5)(0,-16){2}{\circle*{3}}
\multiput(27.5,11.5)(0,-6){3}{\line(0,-1){4}}
\end{picture} }
\newcommand{\bldsh}{
\begin{picture}(34,10)
\put(9,3.5){\circle{11}}
\multiput(8.8,-1.6)(0.5,0.5){12}{\circle*{1}}
\multiput(9.2,8.6)(-0.5,-0.5){12}{\circle*{1}}
\multiput(6.2,-1)(0.5,0.5){15}{\circle*{1}}
\multiput(11.8,8)(-0.5,-0.5){15}{\circle*{1}}
\multiput(14.5,3.5)(6,0){3}{\line(1,0){4}}
\end{picture} }
\newcommand{\half}{
\begin{picture}(36,10)
\put(15,3){\line(-5,2){10}}
\put(15,3){\line(-5,-2){10}}
\put(15,3){\circle*{2}}
\multiput(15,3)(6,0){3}{\line(1,0){4}}
\end{picture} }
\newcommand{\blmix}{
\begin{picture}(50,10)
\put(25,3.5){\circle{11}}
\multiput(24.8,-1.6)(0.5,0.5){12}{\circle*{1}}
\multiput(25.2,8.6)(-0.5,-0.5){12}{\circle*{1}}
\multiput(22.2,-1)(0.5,0.5){15}{\circle*{1}}
\multiput(27.8,8)(-0.5,-0.5){15}{\circle*{1}}
\put(19.5,3.5){\line(-1,0){16}}
\multiput(30.5,3.5)(6,0){3}{\line(1,0){4}}
\end{picture} }
\newcommand{\chmix}{
\begin{picture}(124.5,10)
\put(25,3.5){\circle{11}}
\multiput(24.8,-1.6)(0.5,0.5){12}{\circle*{1}}
\multiput(25.2,8.6)(-0.5,-0.5){12}{\circle*{1}}
\multiput(22.2,-1)(0.5,0.5){15}{\circle*{1}}
\multiput(27.8,8)(-0.5,-0.5){15}{\circle*{1}}
\put(19.5,3.5){\line(-1,0){16}}
\multiput(30.5,3.5)(6,0){3}{\line(1,0){4}}
\put(52,3.5){\circle{11}}
\multiput(51.8,-1.6)(0.5,0.5){12}{\circle*{1}}
\multiput(52.2,8.6)(-0.5,-0.5){12}{\circle*{1}}
\multiput(49.2,-1)(0.5,0.5){15}{\circle*{1}}
\multiput(54.8,8)(-0.5,-0.5){15}{\circle*{1}}
\multiput(57.5,3.5)(6,0){2}{\line(1,0){4}}
\put(71,2.5){...}
\multiput(84,3.5)(6,0){2}{\line(1,0){4}}
\put(99.5,3.5){\circle{11}}
\multiput(99.3,-1.6)(0.5,0.5){12}{\circle*{1}}
\multiput(99.7,8.6)(-0.5,-0.5){12}{\circle*{1}}
\multiput(96.7,-1)(0.5,0.5){15}{\circle*{1}}
\multiput(102.3,8)(-0.5,-0.5){15}{\circle*{1}}
\put(105,3.5){\line(1,0){16}}
\end{picture} }
\newcommand{\cyctwo}{
\begin{picture}(36,19)
\put(9,7){\circle{11}}
\multiput(8.8,1.9)(0.5,0.5){12}{\circle*{1}}
\multiput(9.2,12.1)(-0.5,-0.5){12}{\circle*{1}}
\multiput(6.2,2.5)(0.5,0.5){15}{\circle*{1}}
\multiput(11.8,11.5)(-0.5,-0.5){15}{\circle*{1}}
\multiput(9,12.5)(0.5,0.5){6}{\circle*{1}}
\multiput(9,1.5)(0.5,-0.5){6}{\circle*{1}}
\multiput(14,16.6)(0.5,0.15){6}{\circle*{1}}
\multiput(14,-2.6)(0.5,-0.15){6}{\circle*{1}}
\multiput(19.5,17.35)(0.5,-0.15){6}{\circle*{1}}
\multiput(19.5,-3.35)(0.5,0.15){6}{\circle*{1}}
\multiput(24.5,15)(0.5,-0.5){6}{\circle*{1}}
\multiput(24.5,-1)(0.5,0.5){6}{\circle*{1}}
\put(27,7){\circle{11}}
\multiput(26.8,1.9)(0.5,0.5){12}{\circle*{1}}
\multiput(27.2,12.1)(-0.5,-0.5){12}{\circle*{1}}
\multiput(24.2,2.5)(0.5,0.5){15}{\circle*{1}}
\multiput(29.8,11.5)(-0.5,-0.5){15}{\circle*{1}}
\end{picture} }
\newcommand{\cycone}{
\begin{picture}(31,19)
\put(9,7){\circle{11}}
\multiput(8.8,1.9)(0.5,0.5){12}{\circle*{1}}
\multiput(9.2,12.1)(-0.5,-0.5){12}{\circle*{1}}
\multiput(6.2,2.5)(0.5,0.5){15}{\circle*{1}}
\multiput(11.8,11.5)(-0.5,-0.5){15}{\circle*{1}}
\multiput(9,12.5)(0.5,0.5){6}{\circle*{1}}
\multiput(9,1.5)(0.5,-0.5){6}{\circle*{1}}
\multiput(14,16.6)(0.5,0.15){6}{\circle*{1}}
\multiput(14,-2.6)(0.5,-0.15){6}{\circle*{1}}
\multiput(19.5,17.35)(0.5,-0.15){6}{\circle*{1}}
\multiput(19.5,-3.35)(0.5,0.15){6}{\circle*{1}}
\multiput(24.5,15)(0.33,-0.413){6}{\circle*{1}}
\multiput(24.5,-1)(0.33,0.413){6}{\circle*{1}}
\multiput(27.25,10.75)(0.086,-0.364){8}{\circle*{1}}
\multiput(27.25,3.25)(0.086,0.364){8}{\circle*{1}}
\end{picture} }
\newcommand{\cyone}{
\begin{picture}(28,19)
\put(14,2){\circle{11}}
\multiput(8.5,2)(11,0){2}{\circle*{3}}
\multiput(19.5,2)(0.5,0.5){6}{\circle*{1}}
\multiput(8.5,2)(-0.5,0.5){6}{\circle*{1}}
\multiput(23.6,7)(0.15,0.5){6}{\circle*{1}}
\multiput(4.4,7)(-0.15,0.5){6}{\circle*{1}}
\multiput(24.35,12.5)(-0.15,0.5){6}{\circle*{1}}
\multiput(3.65,12.5)(0.15,0.5){6}{\circle*{1}}
\multiput(22,17.5)(-0.413,0.37){6}{\circle*{1}}
\multiput(6,17.5)(0.413,0.37){6}{\circle*{1}}
\multiput(17.75,20.6)(-0.364,0.1){8}{\circle*{1}}
\multiput(10.25,20.6)(0.364,0.1){8}{\circle*{1}}
\end{picture} }
\newcommand{\cytwo}{
\begin{picture}(28,26)
\put(14,2){\circle{11}}
\multiput(8.5,2)(11,0){2}{\circle*{3}}
\multiput(8.5,20)(11,0){2}{\circle*{3}}
\multiput(19.5,2)(0.5,0.5){6}{\circle*{1}}
\multiput(8.5,2)(-0.5,0.5){6}{\circle*{1}}
\multiput(23.6,7)(0.15,0.5){6}{\circle*{1}}
\multiput(4.4,7)(-0.15,0.5){6}{\circle*{1}}
\multiput(24.35,12.5)(-0.15,0.5){6}{\circle*{1}}
\multiput(3.65,12.5)(0.15,0.5){6}{\circle*{1}}
\multiput(22,17.5)(-0.5,0.5){6}{\circle*{1}}
\multiput(6,17.5)(0.5,0.5){6}{\circle*{1}}
\put(14,20){\circle{11}}
\end{picture} }
\newcommand{\cythr}{
\begin{picture}(42,27)
\multiput(10,2)(22,0){2}{\circle{11}}
\put(21,21.05){\circle{11}}
\put(13.889,-1.889){\circle*{3}}
\put(28.111,-1.889){\circle*{3}}
\put(8.716,7.348){\circle*{3}}
\put(33.284,7.348){\circle*{3}}
\put(15.652,19.766){\circle*{3}}
\put(26.348,19.766){\circle*{3}}
\multiput(13.889,-1.889)(0.5,-0.3){7}{\circle*{1}}
\multiput(28.111,-1.889)(-0.5,-0.3){7}{\circle*{1}}
\multiput(19.75,-4.5)(0.5,0){6}{\circle*{1}}
\multiput(8.716,7.348)(0,0.3){11}{\circle*{1}}
\multiput(33.284,7.348)(0,0.3){11}{\circle*{1}}
\multiput(15.652,19.766)(-0.5,-0.3){7}{\circle*{1}}
\multiput(26.348,19.766)(0.5,-0.3){7}{\circle*{1}}
\multiput(9.2,13.2)(0.25,0.5){6}{\circle*{1}}
\multiput(32.8,13.2)(-0.25,0.5){6}{\circle*{1}}
\end{picture} }
\newcommand{\wlq}{
\begin{picture}(50,12)
\multiput(7,0)(8,0){5}{\oval(4,4)[t]}
\multiput(11,0)(8,0){5}{\oval(4,4)[b]}
\put(5,5.5){\footnotesize {\bf q}}
\put(37,5.5){\footnotesize -{\bf q}}
\end{picture} }
\newcommand{\sixv}{
\begin{picture}(52,10)
\put(15,4){\line(-2,1){10}}
\put(15,4){\line(-1,0){10}}
\put(15,4){\line(-2,-1){10}}
\multiput(15,4)(22,0){2}{\circle*{3}}
\multiput(15,4)(6,0){4}{\line(1,0){4}}
\put(37,4){\line(2,1){10}}
\put(37,4){\line(1,0){10}}
\put(37,4){\line(2,-1){10}}
\end{picture} }

\newcommand{\blo}{
\begin{picture}(39,12)
\thicklines
\put(19.5,-0.4){\circle*{3}}
\put(19.5,-1){\line(-1,0){16}}
\multiput(19.5,-1)(6,0){3}{\line(1,0){4}}
\put(19.5,5.7){\circle{11}}
\end{picture} }
\newcommand{\bloo}{
\begin{picture}(55,12)
\thicklines
\multiput(19.5,-0.4)(16,0){2}{\circle*{3}}
\put(19.5,-1){\line(-1,0){16}}
\put(35.5,-1){\line(1,0){16}}
\multiput(19.5,-1)(6,0){3}{\line(1,0){4}}
\multiput(19.5,5.7)(16,0){2}{\circle{11}}
\end{picture} }
\newcommand{\bllo}{
\begin{picture}(55,12)
\thicklines
\multiput(19.5,-0.4)(16,0){2}{\circle*{3}}
\multiput(19.5,-1)(-6,0){3}{\line(-1,0){4}}
\multiput(35.5,-1)(6,0){3}{\line(1,0){4}}
\put(19.5,-1){\line(1,0){16}}
\multiput(19.5,5.7)(16,0){2}{\circle{11}}
\end{picture} }

\newcommand{\wwv}{
\begin{picture}(54,10)
\put(15,3){\line(-5,2){10}}
\put(15,3){\line(-5,-2){10}}
\multiput(15,3)(24,0){2}{\circle*{2}}
\put(39,3){\line(5,2){10}}
\put(39,3){\line(5,-2){10}}
\multiput(17,3)(8,0){3}{\oval(4,4)[t]}
\multiput(21,3)(8,0){3}{\oval(4,4)[b]}
\end{picture} }
\newcommand{\onebl}{
\begin{picture}(36.22,12)
\put(18.11,4){\circle{11}}
\multiput(17.91,-1.1)(0.5,0.5){12}{\circle*{1}}
\multiput(18.31,9.1)(-0.5,-0.5){12}{\circle*{1}}
\multiput(15.31,-0.5)(0.5,0.5){15}{\circle*{1}}
\multiput(20.91,8.5)(-0.5,-0.5){15}{\circle*{1}}
\put(13.5,7){\line(-3,1){10}}
\put(13.5,1){\line(-3,-1){10}}
\put(22.72,7){\line(3,1){10}}
\put(22.72,1){\line(3,-1){10}}
\end{picture} }
\newcommand{\twobl}{
\begin{picture}(55.44,12)
\multiput(18.11,4)(19.22,0){2}{\circle{11}}
\multiput(17.91,-1.1)(0.5,0.5){12}{\circle*{1}}
\multiput(18.31,9.1)(-0.5,-0.5){12}{\circle*{1}}
\multiput(15.31,-0.5)(0.5,0.5){15}{\circle*{1}}
\multiput(20.91,8.5)(-0.5,-0.5){15}{\circle*{1}}
\multiput(37.13,-1.1)(0.5,0.5){12}{\circle*{1}}
\multiput(37.53,9.1)(-0.5,-0.5){12}{\circle*{1}}
\multiput(34.53,-0.5)(0.5,0.5){15}{\circle*{1}}
\multiput(40.13,8.5)(-0.5,-0.5){15}{\circle*{1}}
\put(13.5,7){\line(-3,1){10}}
\put(13.5,1){\line(-3,-1){10}}
\put(22.72,7){\line(1,0){10}}
\put(22.72,1){\line(1,0){10}}
\put(41.94,7){\line(3,1){10}}
\put(41.94,1){\line(3,-1){10}}
\end{picture} }
\newcommand{\twowbl}{
\begin{picture}(55.44,12)
\multiput(18.11,4)(19.22,0){2}{\circle{11}}
\multiput(17.91,-1.1)(0.5,0.5){12}{\circle*{1}}
\multiput(18.31,9.1)(-0.5,-0.5){12}{\circle*{1}}
\multiput(15.31,-0.5)(0.5,0.5){15}{\circle*{1}}
\multiput(20.91,8.5)(-0.5,-0.5){15}{\circle*{1}}
\multiput(37.13,-1.1)(0.5,0.5){12}{\circle*{1}}
\multiput(37.53,9.1)(-0.5,-0.5){12}{\circle*{1}}
\multiput(34.53,-0.5)(0.5,0.5){15}{\circle*{1}}
\multiput(40.13,8.5)(-0.5,-0.5){15}{\circle*{1}}
\put(13.5,7){\line(-3,1){10}}
\put(13.5,1){\line(-3,-1){10}}
\multiput(23.72,8.15)(8,0){2}{\oval(4,4)[t]}
\put(27.72,8.55){\oval(4,4)[b]}
\multiput(23.72,-0.15)(8,0){2}{\oval(4,4)[b]}
\put(27.72,-0.55){\oval(4,4)[t]}
\put(41.94,7){\line(3,1){10}}
\put(41.94,1){\line(3,-1){10}}
\end{picture} }
\newcommand{\chabl}{
\begin{picture}(97.16,12)
\multiput(18.11,4)(19.22,0){2}{\circle{11}}
\multiput(17.91,-1.1)(0.5,0.5){12}{\circle*{1}}
\multiput(18.31,9.1)(-0.5,-0.5){12}{\circle*{1}}
\multiput(15.31,-0.5)(0.5,0.5){15}{\circle*{1}}
\multiput(20.91,8.5)(-0.5,-0.5){15}{\circle*{1}}
\multiput(37.13,-1.1)(0.5,0.5){12}{\circle*{1}}
\multiput(37.53,9.1)(-0.5,-0.5){12}{\circle*{1}}
\multiput(34.53,-0.5)(0.5,0.5){15}{\circle*{1}}
\multiput(40.13,8.5)(-0.5,-0.5){15}{\circle*{1}}
\put(79.05,4){\circle{11}}
\multiput(78.85,-1.1)(0.5,0.5){12}{\circle*{1}}
\multiput(79.25,9.1)(-0.5,-0.5){12}{\circle*{1}}
\multiput(76.25,-0.5)(0.5,0.5){15}{\circle*{1}}
\multiput(81.85,8.5)(-0.5,-0.5){15}{\circle*{1}}
\put(13.5,7){\line(-3,1){10}}
\put(13.5,1){\line(-3,-1){10}}
\put(22.72,7){\line(1,0){10}}
\put(22.72,1){\line(1,0){10}}
\put(41.94,7){\line(1,0){8}}
\put(41.94,1){\line(1,0){8}}
\put(53.44,3){...}
\put(66.44,7){\line(1,0){8}}
\put(66.44,1){\line(1,0){8}}
\put(83.66,7){\line(3,1){10}}
\put(83.66,1){\line(3,-1){10}}
\end{picture} }
\newcommand{\thrblo}{
\begin{picture}(53.22,26)
\multiput(18.11,4)(17,0){2}{\circle{11}}
\multiput(17.91,-1.1)(0.5,0.5){12}{\circle*{1}}
\multiput(18.31,9.1)(-0.5,-0.5){12}{\circle*{1}}
\multiput(15.31,-0.5)(0.5,0.5){15}{\circle*{1}}
\multiput(20.91,8.5)(-0.5,-0.5){15}{\circle*{1}}
\multiput(34.91,-1.1)(0.5,0.5){12}{\circle*{1}}
\multiput(35.31,9.1)(-0.5,-0.5){12}{\circle*{1}}
\multiput(32.31,-0.5)(0.5,0.5){15}{\circle*{1}}
\multiput(37.91,8.5)(-0.5,-0.5){15}{\circle*{1}}
\put(13.5,7){\line(-3,1){10}}
\put(13.5,1){\line(-3,-1){10}}
\put(23.61,4){\line(1,0){6}}
\put(39.72,7){\line(3,1){10}}
\put(39.72,1){\line(3,-1){10}}
\put(26.61,18.72){\circle{11}}
\multiput(26.41,13.62)(0.5,0.5){12}{\circle*{1}}
\multiput(26.81,23.82)(-0.5,-0.5){12}{\circle*{1}}
\multiput(23.81,14.22)(0.5,0.5){15}{\circle*{1}}
\multiput(29.41,23.22)(-0.5,-0.5){15}{\circle*{1}}
\bezier{50}(20.86,8.763)(22.36,11.36)(23.86,13.96)
\bezier{50}(32.36,8.763)(30.86,11.36)(29.36,13.96)
\end{picture} }
\begin{document}

\title{
\textbf{Critical exponents predicted by grouping of Feynman
diagrams in $\varphi^4$ model}
}

\author{J. Kaupu\v{z}s
\thanks{E--mail: \texttt{kaupuzs@latnet.lv}} \\
Institute of Mathematics and Computer Science, University of Latvia\\
29 Rainja Boulevard, LV--1459 Riga, Latvia}

\date{\today}

\maketitle

\begin{abstract}
Different perturbation theory treatments of the Ginzburg--Landau phase
transition model are discussed. This includes a criticism of the
perturbative renormalization group (RG) approach and a proposal of
a novel method providing critical exponents consistent with the
known exact solutions in two dimensions. The usual perturbation
theory is reorganized by appropriate grouping of Feynman diagrams
of $\varphi^4$ model with $O(n)$ symmetry.
As a result, equations for calculation of the two--point correlation
function are obtained which allow to predict possible exact values of
critical exponents in two and three dimensions
by proving relevant scaling properties of
the asymptotic solution at (and near) the criticality.
The new values of critical exponents are discussed and compared to
the results of numerical simulations and experiments.
\end{abstract}


\section{Introduction}
Phase transitions and critical phenomena is one of the most widely
investigated topics in modern physics. Nevertheless, a limited
number of exact and rigorous results is available~\cite{Baxter}.
Our purpose is to give a critical analysis of the conventional approach
in calculation of critical exponents based on the perturbative
renormalization group (RG) theory~\cite{WF,Ma,Justin} and to propose
a new method which provides results consistent with the known exact solutions.
The usual RG theory is based on several assumptions which could seem to
be plausible since the predicted values of critical exponents are well
confirmed by some numerical results, particularly, by the estimations of
the high--temperature series expansion~\cite{GE,BC}.
The basic hypothesis of RG theory is the existence of
a certain fixed point for the RG transformation.
 The usual RG theory treatment of the Ginzburg--Landau model
is based on the diagrammatic perturbation theory (Feynman diagrams).
Straightforward application of the perturbation theory near
criticality appears to be problematic in the case of the spatial
dimensionality $d<4$ because of the infrared
(i.~e., small wave vector ${\bf k}$) divergence of the expansion terms.
Wilson and Fisher~\cite{WF,Wilson} have proposed a way to overcome this
difficulty by expanding the Feynman diagrams of renormalized perturbation
theory in double series of
$\epsilon=4-d$ and $\ln k$ (as regards just the critical surface).
From this the famous $\epsilon$--expansion of critical exponents
has originated. The first results have been obtained
in~\cite{Wilson,BGZN,BWW}.
Nowadays, an explicit $\epsilon$--expansion
is available up to the fifth order~\cite{VTK,CKT,GLT,KNSCL}.
The $1/n$ expansion of critical exponents~\cite{Ma,Justin,ILF} is based on a
similar idea with the only essential difference that $1/n$ appears as an
expansion parameter (at large $n$) instead of $\epsilon$.
Alternatively, it has been proposed~\cite{Parisi} to expand
the critical exponents in terms of the renormalized coupling constant
at a fixed dimension $d=3$. Later this method has been developed
by several authors~\cite{BNGM,BNM,GJ77,GJ80,GuiJ,AS}.
Apart from the fundamental questions
concerning the validity of the formal expansion in terms of $\ln k$
($\ln k$ diverges at $k \to 0$ !) and similar formal expansions
which lie in the basis of the theory,
a common problem for all these methods is that the resulting series for
critical exponents are divergent (asymptotic), therefore, much efforts have
been devoted to develop appropriate resummation
techniques~\cite{GJ77,GJ80,EMS,FO,EE}.

  In spite of the claims about very accurate
values of critical exponents predicted by the usual RG theory, we
have revealed some serious problems concerning the validity of the
basic assumptions of this theory. In particular, we have demonstrated
that the standard RG treatment is
contradictory and therefore cannot give correct values of critical
exponents. Namely, based on a method which is mathematically correct
and well justified in view of the conventional RG theory, we prove
the nonexistence of the non--Gaussian fixed point predicted by this theory
(Sect.~\ref{sec:RG}). In Sect.~\ref{sec:random} we prove that a correctly
treated diagram expansion provides results which essentially
differ from those of the perturbative (diagrammatic) RG theory.

 Thus, it is worthwhile to search for some alternative analytical methods.
One of such candidates could be the conformal field theory applied to
three--dimensional systems~\cite{WJ}. Note that in two dimensions
this method allows to find the exact critical exponents, and even
calculate the universal ratios of amplitudes. Some
simple, but quite plausible models, like the fractal model of
critical singularity proposed by Tseskis~\cite{Tseskis}, also are
interesting. We have proposed a novel analytical method of
determination of critical exponents in the Ginzburg--Landau model
(Sec.~\ref{sec:my},~\ref{sec:myexp}), and have compared the
predicted exact values of critical exponents to the results of
numerical and real experiments (Sec.~\ref{sec:compare}).

\section{Critical analysis of the perturbative RG method} \label{sec:RG}

Here we consider the Ginzburg--Landau phase transition model within
the usual renormalization group approach to show that this approach
is contradictory. The Hamiltonian of this model in the Fourier
representation reads
\begin{equation} \label{eq:H}
\frac{H}{T}= \sum\limits_{\bf k} \left( r_0+c \,{\bf k}^2 \right)
{\mid \varphi_{\bf k} \mid}^2 + uV^{-1}
\sum\limits_{{\bf k}_1,{\bf k}_2,{\bf k}_3}
\varphi_{{\bf k}_1} \varphi_{{\bf k}_2} \varphi_{{\bf k}_3}
\varphi_{-{\bf k}_1-{\bf k}_2-{\bf k}_3} \;,
\end{equation}
where $\varphi_{\bf k}=V^{-1/2} \int \varphi({\bf x})
\exp(-i {\bf kx}) \, d{\bf x}$ are Fourier components of the scalar
order parameter field $\varphi({\bf x})$, $T$ is the temperature, and
$V$ is the volume of the system. In the RG field theory~\cite{Ma,Justin}
Hamiltonian (\ref{eq:H}) is renormalized by integration of
$\exp(-H/T)$ over $\varphi_{\bf k}$ with $\Lambda/s<k<\Lambda$,
followed by a certain rescaling procedure providing a Hamiltonian
corresponding to the initial values of $V$ and $\Lambda$, where
$\Lambda$ is the upper cutoff of the
$\varphi^4$ interaction. Due to this procedure, additional terms
appear in the Hamiltonian (\ref{eq:H}), so that in general the
renormalized Hamiltonian contains a continuum of parameters.
The basic hypothesis of the RG theory in $d<4$ dimensions is the
existence of a non--Gaussian fixed point $\mu=\mu^*$ for the RG
transformation $R_s$ defined in the space of Hamiltonian parameters, i.e.,
\begin{equation} \label{eq:fixp}
R_s \mu^* = \mu^* \;.
\end{equation}
The fixed--point values of the Hamiltonian parameters are marked by an
asterisk ($r_0^*$, $c^*$, and $u^*$, in particular). Note that
$\mu^*$ is unambiguously defined by fixing the values of $c^*$
and $\Lambda$. According to the RG
theory, the main terms in the renormalized Hamiltonian in
$d=4-\epsilon$ dimensions are those contained in (\ref{eq:H}) with
$r_0^*$ and $u^*$ of the order $\epsilon$,
whereas the additional terms are small corrections of order $\epsilon^2$.

Consider the Fourier transform $G({\bf k}, \mu)$ of the two--point
correlation (Green's) function, corresponding to a point $\mu$.
Under the RG transformation $R_s$ this function transforms as
follows~\cite{Ma}
\begin{equation} \label{eq:RGt}
G({\bf k}, \mu)=s^{2- \eta} \, G(s {\bf k}, R_s \mu) \;.
\end{equation}
Let $G({\bf k}, \mu) \equiv G(k,\mu)$ \,
(at ${\bf k} \ne {\bf0}$ and $V \to \infty$)
be defined within $k \le \Lambda$.
Since Eq.~(\ref{eq:RGt}) holds for any
$s>1$, we can set $s= \Lambda/k$, which at $\mu = \mu^*$ yields
\begin{equation} \label{eq:asyfix}
G({\bf k},\mu^*) = a \, k^{-2 + \eta} \hspace{3ex}
\mbox{for} \;\; k<\Lambda \;,
\end{equation}
where $a= \Lambda^{2- \eta} G(\Lambda, \mu^*)$ is the amplitude
and $\eta$ is the universal critical exponent. According to the
universality hypothesis, the infrared behavior of the Green's
function is described by the same universal value of $\eta$ at
any $\mu$ on the critical surface (with the only requirement that
all parameters of Hamiltonian (\ref{eq:H}) are present), i.e.,
\begin{equation} \label{eq:asy}
G({\bf k},\mu)= b(\mu) \, k^{-2+\eta} \hspace*{3ex}
\mbox{at} \;\; k \to 0\;,
\end{equation}
where
\begin{equation} \label{eq:lim}
b(\mu)=\lim\limits_{k \to 0} k^{2-\eta} \, G({\bf k},\mu) \;.
\end{equation}
According to Eq.~(\ref{eq:RGt}), which holds for any $s=s(k)>1$
and for $s=\Lambda/k$ in particular,
Eq.~(\ref{eq:lim}) reduces to
\begin{equation} \label{eq:b}
b(\mu)=\lim\limits_{k \to 0} k^{2-\eta} s(k)^{2-\eta} \,
G(s {\bf k},R_s \mu) =a \;,
\end{equation}
if the fixed point
$\mu^* = \lim\limits_{s \to \infty} R_s \mu$ exists.
Let us define the function $X({\bf k},\mu)$
as $X({\bf k},\mu) \, = \, k^{-2} G^{-1} ({\bf k},\mu)$.
According to Eqs.~(\ref{eq:asyfix}), (\ref{eq:asy}), and
(\ref{eq:b}), we have (for $k< \Lambda$)
\begin{equation} \label{eq:Xfix}
X({\bf k},\mu^*)= \frac{1}{a} \, k^{-\eta}
\end{equation}
and
\begin{equation} \label{eq:Xcrit}
X({\bf k},\mu)= \frac{1}{a} \, k^{-\eta} \,
+ \, \delta X({\bf k},\mu) \;,
\end{equation}
where $\mu$ belongs to the critical surface,
and $\delta X({\bf k},\mu)$ denotes the correction--to--scaling term.
From (\ref{eq:Xfix}) and (\ref{eq:Xcrit}) we obtain the equation
\begin{equation} \label{eq:eq}
\delta X({\bf k},\mu^* + \delta \mu) =
X({\bf k}, \mu^* + \delta \mu) - X({\bf k}, \mu^*) \; ,
\end{equation}
where $\delta \mu= \mu - \mu^*$.
Since Eq.~(\ref{eq:eq}) is true for any small deviation $\delta \mu$
satisfying the relation
\begin{equation} \label{eq:ff}
\mu^* =\lim\limits_{s \to \infty} R_s(\mu^* +\delta\mu) \;,
\end{equation}
we choose $\delta \mu$ such that $\mu^* \Rightarrow \mu^* +
\delta\mu$ corresponds to the variation of the Hamiltonian parameters
 $r_0^* \Rightarrow r_0^* + \delta r_0$,
$c^* \Rightarrow c^* + \delta c$, and
$u^* \Rightarrow u^* + \,\epsilon \times \Delta$,
where $\Delta$ is a small constant.
The values of $\delta r_0$ and $\delta c$ are choosen to fit the
critical surface and to meet the condition (\ref{eq:ff}) at fixed
$c^*=1$ and $\Lambda=1$. In particular, quantity $\delta c$ is found
$\delta c \,=\, B \; \epsilon^2 \, +o(\epsilon^3)$
with some (small) coefficient $B=B(\Delta)$, to compensate the shift in
$c$ of the order $\epsilon^2$ due to the renormalization (cf.~\cite{Ma}).
The formal $\epsilon$--expansion of $\delta X({\bf k}, \mu)$,
defined by Eq.~(\ref{eq:eq}), can be obtained in the usual way from the
perturbation theory. This yields
\begin{equation} \label{eq:expas}
\delta X({\bf k}, \mu) = \epsilon^2 \, [\, C_1(\Delta) +
C_2(\Delta) \, \ln k \, ] \, +o(\epsilon^3) \hspace*{3ex}
\mbox{at} \;\; k \to 0 \;,
\end{equation}
where $C_1(\Delta)$ and $C_2(\Delta)$ ($C_2 \ne 0$) are coefficients
independent on $\epsilon$.

It is commonly accepted in the RG field theory to make an
expansion like (\ref{eq:expas}), obtained from the diagrammatic
perturbation theory, to fit an asymptotic expansion in $k$ powers,
thus determining the critical exponents.
In general, such a method is not rigorous since,
obviously, there exist such functions which do not contribute
to the asymptotic expansion in $k$ powers at $k \to 0$, but give a
contribution to the formal $\epsilon$--expansion at any fixed
$k$. Besides, the expansion coefficients do not vanish at $k \to 0$.
Trivial examples of such functions are
$\epsilon^m \, \exp(-\epsilon k^{-\epsilon})$
and $\epsilon^m \,[1-\tanh(\epsilon \,k^{-\epsilon})]$ where $m$ is integer.
Nevertheless, according to the general ideas of the RG theory
(not based on Eq.~(\ref{eq:eq})),
in the vicinity of the fixed point the asymptotic expansion
\begin{equation} \label{eq:Xas}
X({\bf k},\mu)= \frac{1}{a} k^{-\eta} +
b_1 k^{\epsilon+o(\epsilon^2)} + b_2 k^{2+o(\epsilon)} + ...
\end{equation}
is valid not only at $k \to 0$, but within $k<\Lambda$. The latter means
that terms of the kind $\epsilon^m \, \exp(-\epsilon k^{-\epsilon})$
are absent or negligible.
Thus, if the fixed point does exist, then we can
obtain correct $\epsilon$--expansion of $\delta X({\bf k},\mu)$
at small $k$ by expanding the term $b_1 k^{\epsilon+o(\epsilon^2)}$
(with $b_1=b_1(\epsilon,\Delta)$) in
Eq.~(\ref{eq:Xas}) in $\epsilon$ powers, and the result must
agree with (\ref{eq:expas}) at small $\Delta$, at least.
The latter, however, is impossible
since Eq.~(\ref{eq:expas}) never agree with
\begin{equation}
\delta X({\bf k},\mu) = b_1(\epsilon,\Delta) \,
\left[\, 1+ \epsilon \ln k +o(\epsilon^2) \, \right]
\end{equation}
obtained from (\ref{eq:Xas}) at $k \to 0$. Thus,
in its very basics the perturbative RG method in $4-\epsilon$ dimensions
is contradictory. From this we can conclude that the initial
assumption about existence of a certain fixed point, predicted
by the RG field theory in $4-\epsilon$ dimensions, is not valid.

\section{A model with quenched randomness} \label{sec:random}

Here we consider the Ginzburg--Landau phase transition model with $O(n)$
symmetry (i.e., the $n$--vector model) which includes a quenched randomness,
i.e., a random temperature disorder. One of the basic ideas of the
perturbative RG theory is that $n$ may be considered as a continuous
parameter and the limit $n \to 0$  makes sense describing the self--avoiding
random walk or statistics of polymers~\cite{Ma,Justin}.
We have proven rigorously that within the diagrammatic perturbation theory
the quenched randomness does not change the critical exponents at $n \to 0$,
which is in contrast to the prediction of the conventional RG theory
formulated by means of the Feynman diagrams.

 The Hamiltonian of the actually considered model is
\begin{eqnarray} \label{eq:Hr}
H/T &=& \int \left[ \left(r_0+ \sqrt{u} \, f({\bf x}) \right)
\varphi^2({\bf x}) + c \, (\nabla \varphi({\bf x}) )^2 \right] d{\bf x}
\\
&+& \, uV^{-1} \sum\limits_{i,j,{\bf k}_1,{\bf k}_2,{\bf k}_3 }
\varphi_i({\bf k}_1) \varphi_i({\bf k}_2) \, u_{{\bf k}_1+{\bf k}_2} \,
\varphi_j({\bf k}_3) \varphi_j(-{\bf k}_1-{\bf k}_2-{\bf k}_3) \nonumber
\end{eqnarray}
which includes a random temperature (or random mass) disorder
represented by the term $\sqrt{u}\,f({\bf x})\,\varphi^2({\bf x})$.
For convenience, we call this model the random model. In Eq.~(\ref{eq:Hr})
$\varphi({\bf x})$ is an $n$--component vector with components
$\varphi_i({\bf x})= \linebreak V^{-1/2} \sum_{k<\Lambda} \varphi_i({\bf k})
e^{i{\bf kx}}$,
depending on the coordinate ${\bf x}$, and
$f({\bf x})=V^{-1/2} \sum_{\bf k} f_{\bf k} e^{i{\bf kx}}$ is a random
variable with the Fourier components
$f_{\bf k}=V^{-1/2} \int f({\bf x}) e^{-i{\bf kx}} d{\bf x}$.
The only allowed configurations of the order parameter field
$\varphi({\bf x})$ are those corresponding to $\varphi_i({\bf k})=0$
at $k>\Lambda$. This is the limiting case $m \to \infty$ of the
model where all configurations are allowed, but Hamiltonian~(\ref{eq:Hr})
is completed by term $\sum_{i,{\bf k}} \left( k/\Lambda \right)^{2m}
\mid \varphi_i({\bf k}) \mid^2$.

The system is characterized by the two--point correlation function
$G_i({\bf k})$ defined by the equation
\begin{equation}
\left< \varphi_i({\bf k}) \varphi_j(-{\bf k}) \right>
= \delta_{i,j} \, G_i({\bf k}) = \delta_{i,j} \, G({\bf k})\;.
\end{equation}
It is supposed that the averaging is performed over the $\varphi({\bf x})$
configurations and then over the $f({\bf x})$ configurations with a fixed
(quenched) Gaussian distribution $P(\{ f_{\bf k} \})$ for the set of Fourier
components $\{ f_{\bf k} \}$,
i.~e., our random model describes a quenched randomness.

We have proven the following theorem.
\vspace*{1ex}

{\it Theorem}. \, In the limit $n \to 0$, the perturbation expansion
of the correlation function  $G({\bf k})$ in  $u$  power series for the
random model with the Hamiltonian (\ref{eq:Hr}) is identical to the
perturbation expansion for the corresponding model with the Hamiltonian
\begin{eqnarray} \label{eq:Hp}
H/T &=& \int \left[ r_0 \, \varphi^2({\bf x})
+ c \, (\nabla \varphi({\bf x}) )^2 \right] d{\bf x} \\
&+& \, uV^{-1} \sum\limits_{i,j,{\bf k}_1,{\bf k}_2,{\bf k}_3 }
\varphi_i({\bf k}_1) \varphi_i({\bf k}_2) \, \tilde u_{{\bf k}_1+{\bf k}_2} \,
\varphi_j({\bf k}_3) \varphi_j(-{\bf k}_1-{\bf k}_2-{\bf k}_3) \nonumber
\end{eqnarray}
where $\tilde u_{\bf k}=u_{\bf k}- {1 \over 2}
\left< {\mid f_{\bf k} \mid}^2 \right>$. \medskip

   For convenience, we call the model without the term
$\sqrt{u} \, f({\bf x}) \, \varphi^2({\bf x})$ the pure model, since this
term simulates the effect of random impurities~\cite{Ma}.
\vspace*{1ex}

 {\it Proof of the theorem}. \, According to the rules of the diagram
technique, the formal expansion for $G({\bf k})$ involves all connected
diagrams with two fixed outer solid lines. In the case of the pure model,
diagrams are constructed of the vertices {\mbox \zigv,} with factor
$-uV^{-1} \tilde u_{\bf k}$ related to any zigzag line with wave vector
${\bf k}$.
The solid lines are related to the correlation function in the Gaussian
approximation $G_0({\bf k})=1/ \left(2r_0+2ck^2 \right)$. Summation over
the components $\varphi_i({\bf k})$ of the vector $\varphi({\bf k})$
yields factor $n$ corresponding to each closed loop of solid lines in the
diagrams. According to this, the formal perturbation expansion is defined
at arbitrary $n$. In the limit $n \to 0$, all diagrams of $G({\bf k})$
vanish except those which do not contain the closed loops. In such a way,
for the pure model we obtain the expansion
\begin{equation} \label{eq:expu}
G({\bf k})= \Gausd + \, \zigd + \, ... \;.
\end{equation}
 In the case of the random model, the diagrams are constructed of
the vertices \dshv and {\mbox \dotv.} The factors
$uV^{-1} \left< {\mid f_{\bf k} \mid}^2 \right>$ correspond to the
coupled dotted lines and the factors $-uV^{-1} u_{\bf k}$ correspond to
the dashed lines. Thus, we have
\begin{equation} \label{eq:exra}
G({\bf k}) = \Gausd + \left[ \dashd + \dotd \right] + \, ... \;.
\end{equation}
In the random model, first the correlation function $G({\bf k})$
is calculated at a fixed $\{ f_{\bf k} \}$
(which corresponds to connected diagrams where solid lines are coupled, but the
dotted lines with factors $-\sqrt{u} \,V^{-1/2} f_{\bf k}$ are not coupled),
performing the averaging with the weight $P( \{ f_{\bf k} \})$ over the
configurations of the random variable
(i.e., the coupling of the dotted lines) afterwards.
According to this procedure, the diagrams of the random model in
general (not only at $n \to 0$) do not contain parts like {\mbox \bldot,}
{\mbox \blddot,} {\mbox \bltdot,} etc.,
which would appear only if unconnected (i.e., consisting of separate
parts) diagrams would be considered before the coupling of dotted lines.

It is evident from
Eqs.~(\ref{eq:expu}) and (\ref{eq:exra}) that all diagrams of the random
model are obtained from those of the pure model if any of the zigzag
lines is replaced either by a dashed or by a dotted line, performing
summation over all such possibilities. Such a method is valid in
the limit $n \to 0$, but not in general. The problem is that, except
the case $n \to 0$, the diagrams of the pure model contain parts like
{\mbox \blzig,} {\mbox \bldzig,} {\mbox \bltzig,} etc. If all the depicted
here zigzag lines are replaced by the dotted lines, then we obtain diagrams
which are not allowed in the random model, as explained
before. At $n \to 0$, the only problem is to determine the
combinatorial factors for the diagrams obtained by the above replacements.
For a diagram constructed of $M_1$ vertices \dshv and $M_2$ vertices \dotv
the combinatorial factor is the number of possible different couplings of
lines, corresponding to the given topological picture, divided by
$M_1 ! M_2 !$.

  Our further consideration is
valid also for the diagrams of free energy (at $n \to 0$
represented by the main terms containing single loop of solid lines)
and of $2m$--point correlation function.
We define that all diagrams which can be obtained from the
$i$--th diagram (i.e., the diagram of the $i$--th topology) of the pure
model, belong to the  $i$--th group. Obviously, all diagrams of the
$i$--th group represent a contribution of order $u^l$, where $l$ is
the total number of vertices \zigv in the $i$--th diagram.
The sum of the diagrams of the
$i$--th group can be found by the following algorithm.
\begin{itemize}
\item[1.]
Depict the $i$--th diagram of pure model in an a priori defined way.
\item[2.] Choose any one replacement
of the vertices \zigv by \dshv and {\mbox \ddotv,} and perform the summation
over all such possibilities. For any specific choice we consider only one
of the equivalent $M_1!M_2!$ distributions of the numbered $M_1$ vertices
\dshv and $M_2$ vertices \dotv over the fixed numbered positions instead of
the summation over all these distributions
with the weight $1/(M_1!M_2!)$. Thus, at this step the combinatorial factor
for any specific diagram is determined as
the number of possible distributions of lines (numbered before coupling)
for one fixed location of vertices consistent with the picture defined
in step 1.
\item[3.] The result of summation in
step 2  is divided by the number of independent symmetry transformations
(including the identical transformation) for the considered $i$--th diagram
constructed of vertices \zigv, since the same (original and transformed)
diagrams were counted as different.
\end{itemize}
   Note that the location of any vertex \dshv is defined by fixing the
position of dashed line, the orientation of which is not fixed. According
to this, the summation over all possible distributions of lines (numbered
before coupling) for one fixed location of vertices
yields factor $8^{M_1}4^{M_2/2}$. The $i$--th
diagram of the pure model also can be calculated by such an algorithm.
In this case we have $8^l$ line distributions, where $l=M_1+M_2/2$\,
is the total number of vertices \zigv in the  $i$--th diagram. Obviously,
the summation of diagrams of the  $i$--th  group can be performed with
factors $8^l$ instead of $8^{M_1}4^{M_2/2}$, but in this case
twice smaller factors
must be related to the coupled dotted lines.
The summation over all possibilities where zigzag lines are replaced by dashed
lines with factors $-uV^{-1}u_{\bf k}$ and by dotted lines with factors
${1 \over 2} uV^{-1} \left< {\mid f_{\bf k} \mid}^2 \right>$, obviously,
yields a factor
$uV^{-1} \left(-u_{\bf k}+{1 \over 2} \left< {\mid f_{\bf k} \mid}^2
\right> \right) \equiv -uV^{-1} \tilde u_{\bf k}$ corresponding to each
zigzag line with wave vector ${\bf k}$. Thus, the sum over the diagrams of
the  $i$--th group is identical to the  $i$--th diagram of the pure model
defined by Eq.~(\ref{eq:Hp}). By this the theorem has proved
not only for the two--point correlation function, but also for
$2m$--point correlation function and free energy.

If, in general, the factor $\sqrt{u}$ in Eq.~(\ref{eq:Hr}) is replaced by
$\sqrt{u'}$, where $u'$ is an independent expansion parameter,
then our analysis leads to the above relation between
diagrams for $u \, \tilde u_{\bf k}= u \, u_{\bf k}- {u'\over 2}
\left< {\mid f_{\bf k} \mid}^2 \right>$.
According to this, at $n \to 0$ the pure
and random models cannot be distinguished within the
diagrammatic perturbation theory. If, in principle, critical
exponents can be determined from the diagram expansions at
$n \to 0$, as it is suggested in the usual RG theory, then the same
critical exponents should be provided for both models at $n \to 0$.
In such a way, we conclude that the RG method is
not correct because the above condition is violated.
As compared to our simple treatment of the random model, the RG
treatment includes additional Feynman diagrams because the
Hamiltonian becomes more complicated after the renormalization.
However, this does not enable to find the
difference between both models: the original
information, when one starts the perturbative renormalization of
Hamiltonian~(\ref{eq:Hr}), is contained in the Feynman diagrams
we considered, but the renormalization by itself does not create
new information about the model. Really, by renormalization we merely
``forget'' some information about the short--wave fluctuations
to make that for the long--wave fluctuations easier accessible.
Thus, our conclusion remains true.

\section{Equations from reorganized perturbation theory} \label{sec:my}

 As we have already discussed in Sect.~\ref{sec:RG}, it is not a
rigorous method to make a formal expansion like (\ref{eq:expas})
and to try calculate the critical exponents therefrom. We
propose another treatment of the diagrammatic perturbation theory
for the Ginzburg--Landau model defined by Eq.~(\ref{eq:Hp}),
where $u \, \tilde u_{\bf k}= u_{\bf k}$.
The basic idea is to obtain suitable equations by appropriate grouping
of the diagrams. Suitable are such equations which allow to find
the asymptotic expansions at the critical point directly in $k$
power series, but not in terms of the formal parameter $\ln k$
(as in Eq.~\ref{eq:expas}) which diverges at $k \to 0$.

\subsection{Some fundamental problems of the perturbation theory}
\label{sec:fp}

One of the problem which can arise in any perturbation theory is that the
perturbation expansion alone does not define unambiguously the original
function. For example, all expansion coefficients in the formal expansion of
$\exp \left(-1/u \right)$ with $u$ considered as an expansion parameter at
$u=+0$  are zero, whereas the original function is not zero.
However, if quantities $A$ and $B$ have the same perturbation expansion in a
power series of $u$, then $A(u)-B(u)=C(u)$ holds, where $C(u)$
is some function of $u$ with all the expansion coefficients
equal to zero.
Accordingly, the diagram technique principally allows, in the worst case
(where $C(u) \ne 0$ ), to find $G({\bf k})$ with an accuracy to some unknown
function which tends to zero at $u \to 0$ faster than $u^l$ at
any $l>0$. We have shown in Sect.~\ref{sec:detex} that true (exact) critical
exponents can be obtained neglecting this function.

    Another problem is that the considered formal diagram expansions diverge.
Nevertheless, the following way is possible, which lies
in the basis of our diagrammatic treatment. First, we build up a perturbation
expansion of quantity $A$ which is necessary to be found. Then we seek such a
quantity $B$ (represented by converging sums and integrals), the perturbation
expansion of which is identical to that we have built up.
According to the above consideration, we
have $A(u)=B(u)+C(u)$, where $C(u)$ is an insignificant correction (or zero).
In this case it is not necessary that the perturbation sum converge if
calculated in a straightforward way.
Various manipulations
with diagram blocks appearing in Sec.~\ref{sec:my} are defined as constructions
of corresponding formal (diverging) expressions which, however, provide correct
expansion in terms of $u$. Various diagram representations of a given
quantity yielding the same expansion in a power series of $u$ are defined as
equivalent.

\subsection{The diagram notation} \label{sec:notation}

Here we define some diagram notations appearing in Sec.~\ref{sec:my}.


\textit{Coupled diagram} is defined as any diagram which does not contain uncoupled lines,
i.~e., any line starts from some kink and ends in the same or another kink, where
"kink" means a merging point of solid and dashed lines of a
vertex {\mbox \dshv.}

\textit{The self--energy block}   \selfek  denotes the perturbation sum involving all
connected diagrams of this kind, i.~e., the sum of all specific self--energy blocks or
diagrams which cannot be reduced to a linear chain like   \selfch
consisting of two or more blocks. The simplest specific self--energy blocks are
\vspace{5pt} \\
$\sked = -2u_{\bf 0} V^{-1} n \sum\limits_{\bf q} G_0({\bf q})$
and
$\skedd = -4 V^{-1} \sum\limits_{\bf q} u_{\bf q} \, G_0({\bf k}-{\bf q})$.
\vspace{10pt} \\
Factors corresponding to the lines marked by crosses are omitted.
Each case of topologically nonequivalent coupling of lines
corresponds to one diagram.

\textit{Skeleton diagram}  is defined as a connected diagram, containing no
parts like \mbox{\selfe,}
with factors $G({\bf k})$ corresponding to the solid lines. For example, the
simplest coupled skeleton diagrams are
\\
$\skd =
 -u_{\bf 0} V^{-1} \left( \sum\limits_{i,{\bf q}} G_i({\bf q}) \right)^2$
and $\skdd =
-2V^{-1} \sum\limits_i \sum\limits_{\bf k} \sum\limits_{\bf q}
u_{\bf k} \, G_i({\bf q}) G_i({\bf k}-{\bf q})$.
Note that index $i$
can be removed, replacing $\sum_i$ with factor $n$.
The simplest skeleton diagrams like \selfek with two outer lines (not coupled
diagrams) are the same as the above given self--energy diagrams, but with
$G({\bf k})$ instead of $G_0({\bf k})$.

\textit{Single block} is defined as a connected diagram with two outer (broken)
dashed lines which cannot be reduced to a linear chain like \ssch
consisting of two or more blocks.

\textit{The single skeleton block}  \ssbkn  denotes the perturbation sum
involving all single blocks that belong to skeleton diagrams, i.~e., the sum
of all specific single skeleton blocks. The outer (broken) dashed lines or
kinks in this case are marked by $1$ and $2$ (in general kinks
are not depicted).
Factor $-V^{-1} u_{\bf k}$ corresponds to the pair of these lines. Each case of
topologically nonequivalent coupling of lines
with respect to fixed kinks 1 and 2, considered as nonequivalent, corresponds
to one diagram of {\mbox \ssbk.} For instance,
\vspace{5pt} \\
$\ssbll = 16 u_{\bf k} V^{-2} \sum\limits_i \sum\limits_{\bf q}
\sum\limits_{\bf p} G_i({\bf q}) G_i({\bf k}-{\bf q}) \,
u_{{\bf q}-{\bf p}} \, G_i({\bf p}) G_i({\bf k}-{\bf p})$.
\vspace{5pt} \\
\textit{Combinatorial factor} corresponding to any specific diagram is not
given explicitly, but is implied in the diagram itself. It can be
calculated following the scheme in Sec.~\ref{sec:random}
(but without the replacements, since now we have only one kind of
vertices).

\subsection{Expansion of $G({\bf k})$ in terms of skeleton diagrams}
\label{sec:exG}

 It is suitably to have a diagram expansion for $G({\bf k})$  where the true
correlation function $G({\bf k})$ is related to solid lines instead of
$G_0({\bf k})$. To obtain this, first let us consider the quantity
$\widetilde \Sigma({\bf k})$  defined by equation
\begin{equation} \label{eq:Sigdef}
G({\bf k})= \frac{G_0({\bf k})}{1 - 2 \widetilde \Sigma({\bf k})
\, G_0({\bf k})} \; .
\end{equation}
It is well known~\cite{Ma} that terms of the perturbation expansion of
$\widetilde \Sigma({\bf k})$ are diagrams of the self--energy
block \mbox{\selfek.} Note only that the self--energy defined in
Ref.~\cite{Ma} corresponds to $-2 \widetilde \Sigma({\bf k})$.

\begin{figure}
\centerline{\psfig{figure=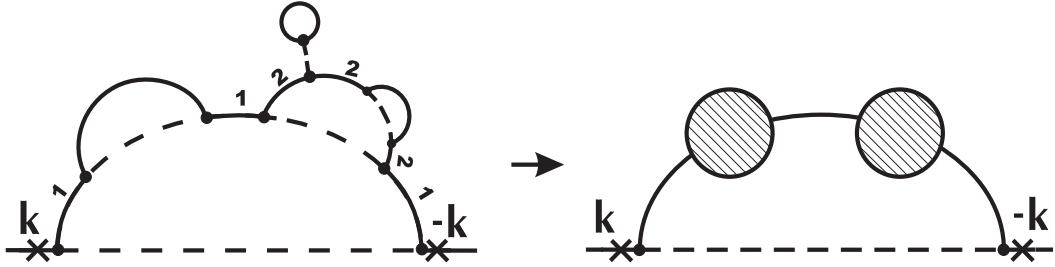,width=14cm,height=3.5cm}}
\caption{\small An example of replacement, where the blocks of  primary chain
inside a diagram are replaced by the self--energy blocks, yielding a diagram of
set~$A$. The connecting solid lines of the primary chain are marked
by~$1$. There is another (not primary) chain,
marked by~$2$, which is built into a block of the primary chain.}
\label{figg1}
\end{figure}

    The desired expansion with $G({\bf k})$ instead of $G_0({\bf k})$ is
obtained by grouping of diagrams involved in the self--energy block.
    First we consider specific self--energy blocks, called the primal diagrams,
from which no block of the kind \selfe can be extracted. Then we consider the
set $A$ of diagrams obtained by extending the solid lines inside the primal
diagrams by adding all possible numbers $m \in [0; \infty]$ of self--energy
blocks to each of these lines. In
particular, the diagram of set  $A$, depicted on the right hand side of
Fig.~\ref{figg1}, is obtained from the primal diagram \skedd by adding two
($m=2$) self--energy blocks to the inner solid line.
Obviously, all diagrams of set $A$ are self--energy diagrams,
i.~e., they cannot be split in two blocks
like \selftwo. Besides, any specific self--energy block is contained
in a perturbation sum represented by one of diagrams of
set $A$. This is proved considering primary chains contained inside the
specific self--energy blocks. A primary chain is defined as a linear chain of
specific self--energy blocks which does not belong to (i.~e., is not built into) a block
of some other linear chain inside the diagram.
If we replace the blocks of
primary chains by {\mbox \selfe,} we obtain a diagram of set $A$, i.~e., a
diagram consisting of separate linear chains of the
new blocks, the initial diagram being involved as a particular case
(since \selfe involves all specific
blocks of such kind). A specific example of such a replacement is illustrated
in Fig.~\ref{figg1}. Thus, set  $A$  contains all specific self--energy blocks.

Because blocks are distributed
independently over separate linear chains, summation over all possible
lengths $N$ of linear chains yields a factor\linebreak
$\sum_N G_0({\bf k}) \left( 2 G_0({\bf k}) \times \selfek \right)^N$
identical to the perturbation sum of $G({\bf k})$ for each of
the original solid lines (in a primal diagram) with wave vector ${\bf k}$.
Thus an equivalent diagram
representation of $\widetilde \Sigma({\bf k})$
(Sect.~\ref{sec:fp}) is obtained if the perturbation sum over all diagrams of
set  $A$ is replaced by the perturbation sum
over the primal diagrams in which $G_0({\bf k})$
is replaced by $G({\bf k})$. According to the definition in
Sect.~\ref{sec:notation}, diagrams of the new expansion are skeleton diagrams
like {\mbox \selfek.}

    Let us assume that there exists a quantity $D(G)$ the perturbation sum of
which consists of all coupled (containing no outer lines) skeleton diagrams.
$D(G)$ have to be considered as a
function of discrete variables $G_i({\bf k})$ (although
$G_i({\bf k})=G({\bf k})$ holds, the cases with different $i$ formally
are nonequivalent) corresponding to the set of discrete wave vectors ${\bf k}$.
All skeleton diagrams of $\widetilde \Sigma({\bf k})$, i.~e., those
like \selfek (and these exclusively) can be obtained by
breaking a line with wave vector ${\bf k}$ in a coupled diagram
of $D(G)$ and removing factor
$G({\bf k})$ corresponding to this line. This procedure is identical to the
derivation of the perturbation sum of $D(G)$
with respect to $G_i({\bf k})$.
 Consequently, for any quantity $D(G)$, having the
diagram expansion represented by coupled skeleton diagrams, the equation
\begin{equation} \label{eq:sig}
\widetilde \Sigma({\bf k}) = \frac{\partial D(G)}
{\partial G_i({\bf k})} - \vartheta({\bf k})
\end{equation}
is true with some function $\vartheta({\bf k})$
providing zero contribution to the expansion in power series of $u$
(cf. Sec.~\ref{sec:fp}).

\subsection{Representation of skeleton diagrams by single skeleton blocks}
\label{sec:rsd}

In this section the skeleton diagrams are represented by single skeleton blocks.
The only skeleton diagram among those represented as \skbb is
{\mbox \skd,} i.~e.,  if the specific block \bldsh  contains more than one solid line, obtained by
coupling of two solid lines of the "half--vertex" {\mbox \half,} then we can
extract the part \selfe  by breaking the two solid lines coupling it to the rest of the block.

 Other skeleton diagrams without outer lines, representing the perturbation sum of
the quantity
\begin{equation} \label{eq:Dg}
D^*(G)=D(G) - \skd
\end{equation}
 can be represented in one way, at least, by cyclically coupled
specific single skeleton blocks. Really,
we can break any of the dashed lines in the skeleton diagram of
$D^*(G)$ to obtain a connected diagram like
{\mbox \ssb.} In this case \ssb denotes any skeleton diagram with two outer
dashed lines. If the obtained diagram cannot be split to make two
{\mbox \ssb \ssb} by interrupting one more
dashed line, we can join the broken dashed line again representing this
as a single skeleton block {\mbox \cycone.} If the obtained
diagram can be split in two such blocks but cannot be split in three by interrupting
one more dashed line, we can represent this by a cycle like
\cyctwo with two specific
single skeleton blocks, when coupling again, and so on.
 Any such cyclical coupling of specific single skeleton blocks
corresponds to one of the skeleton diagrams because it is impossible to extract any
block of the kind \selfe therefrom. It cannot be extracted from any specific
single skeleton block according to the definition, as well as in any other
way because subblocks of the kind \blmix
(constructed of odd number of solid lines by coupling) are principally
impossible (in the opposite case parts like \chmix could be extracted).

\subsection{Summation of the simplest skeleton diagrams}
\label{sec:simpled}

   Consider now a contribution to $D(G)$ (denoted by $D^{(0)}(G)$) of diagrams
containing the simplest specific single skeleton blocks
\begin{equation} \label{eq:simpd}
\Sigma^{(0)}({\bf k})= \ssbl = -2 u_{\bf k} V^{-1}
\sum\limits_{i,{\bf q}} G_i({\bf q}) G_i({\bf k}-{\bf q})
\end{equation}
exclusively. This contribution is not merely a formal sum since
at small $u_{\bf k}$ (at $r_0>0$) it can be obtained by straightforward
summation, i.~e.,
\begin{eqnarray}
D^{(0)}(G) &=& \skd + \cyone + \cytwo + \cythr + ... \nonumber
\vspace*{15pt} \\
&=& \skd - {1 \over 2} \sum\limits_{\bf q} \ln
\left[ 1- 2 \Sigma^{(0)}({\bf q}) \right] \; . \label{eq:D0}
\end{eqnarray}
Equation~(\ref{eq:D0}) is true in any case if $D^{(0)}(G)$ is defined as a quantity having this
diagram expansion. The sum~(\ref{eq:D0}) is calculated by the following method. To calculate
the combinatorial factor for a cycle comprised of $N$ blocks, first we count all
possible distributions of $N$  numbered vertices over $N$ sequentially numbered fixed
sites along the cycle with all possible  distributions of  $4N$  uncoupled lines, which are
then coupled in $2N$ lines. Then, the result is corrected taking into account that
couplings obtained in such a way contain equivalent ones which differ merely by a
diagram having been rotated as a whole or (and) transformed by a mirror--symmetry
transformation, as well as in $2^N$ ways transformed by a mirror--symmetry
transformation of each \ssbe  block separately. $2N \cdot 2^N$  such independent
transformations exist.

\subsection{Grouping of the skeleton diagrams}

In this section we reorganize the perturbation expansion 
which is necessary for calculation of $D^*(G)$.

\begin{figure}
\centerline{\psfig{figure=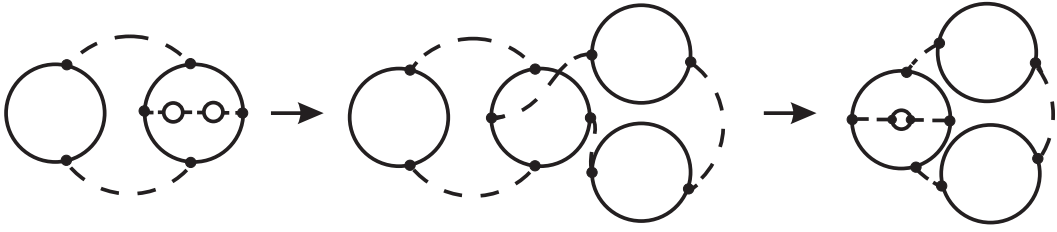,width=14cm,height=3cm}}
\caption{\small An example of transformation leading to a different representation
of the diagram.}
\label{figg2}
\end{figure}

First, we have got a formal diagram equation performing
manipulations similar to those in
Sect.~\ref{sec:simpled}, but including all diagrams of the single skeleton block.
  A problem arises with more complicated diagrams than those considered in
Sect.~\ref{sec:simpled} because of additional symmetry related to different
representations of a given diagram
by specific cyclically coupled single skeleton blocks. In
Fig.~\ref{figg2}, an example of such a
transformation is shown, which leads to a different representation of the diagram,
retaining the couplings of lines unchanged.
Two representations of a given coupled
diagram with numbered vertices are considered as different or nonequivalent if they
cannot be transformed into each other without using a transformation of this kind.

  If we now consider the perturbation sum of cyclically coupled
blocks \ssb with the combinatorial factors calculated as in
Sect.~\ref{sec:simpled}
without account for the above considered additional symmetry, then a specific diagram of
the $i$--th topology (or the $i$--th skeleton diagram) in the perturbation sum will
have weight $m_i$, where $m_i$
is the number of possible different representations of this diagram by specific
cyclically coupled single skeleton blocks. This leads to the following formal identity
of perturbation sums
\begin{equation} \label{eq:formal}
-{1 \over 2} \sum\limits_{\bf q} \ln \left[ 1-2 \times \ssbq
\right] = \sum\limits_i^{\hspace*{3ex} \prime} m_i C_i \; ,
\end{equation}
where $C_i$ denotes the $i$--th coupled skeleton diagram, the
apostrophe " $\prime$ " showing that diagram  \skd  is omitted.

    Number $m_i$  can be identified with the number of linear chains consisting of
dashed lines and blocks of the kind \ssb (or a single dashed line) contained
within diagram $C_i$, counting all such chains contained inside specific single skeleton
blocks that comprise the cycle in a given representation and the chain representing the
cycle itself. It holds because any linear chain contained in a single skeleton block can
represent a cycle in one of the new representations of the diagram obtained by such a
transformation as illustrated in Fig.~\ref{figg2}. Note that linear chains can be built into
each other in all possible ways (like in a fractal), and all these chains
(of all levels) are counted as different.

We have avoided explicit counting of linear chains as follows.
We have proven the following: If there exists a quantity $\Sigma({\bf q},\zeta)$ which has the
perturbation expansion $\Sigma({\bf q},\zeta)=\sum\limits_i
\Sigma_i ({\bf q}) \, \zeta^{n_i}$, where $\Sigma_i ({\bf q})$ is a specific
single skeleton block of the $i$--th
topology and $n_i$ is the number of linear chains contained in this block,
and if there is a solution of the equation
\begin{equation} \label{eq:D}
D^*(G,\zeta)= -{1 \over 2} \sum\limits_{\bf q} \ln
\left[ 1 - 2 \Sigma({\bf q},\zeta) \right]
- \zeta \frac{\partial}{\partial \zeta} D^*(G,\zeta)
\end{equation}
with the boundary condition
\begin{equation} \label{eq:Db}
D^*(G,0)= -{1 \over 2} \sum\limits_{\bf q}
\ln \left[ 1 - 2 \Sigma^{(0)}({\bf q}) \right] \; ,
\end{equation}
then quantity $D^*(G,\zeta)$, calculated from this equation, has the
perturbation expansion
$D^*(G,\zeta)=\sum\limits_i^{\hspace*{2ex} \prime} C_i \, \zeta^{m_i-1}$.
Since Eqs.~(\ref{eq:D}) and (\ref{eq:Db}) define the solution unambiguously,
it is necessary only to prove the
identity of perturbation expansions in the case if $D^*(D,\zeta)$ has the
expansion $\sum\limits_i^{\hspace*{2ex} \prime} C_i \, \zeta^{m_i-1}$.
The identity of diagram expansions in~(\ref{eq:Db}) corresponding
to $m_i=1$ follows directly from the consideration in
Sect.~\ref{sec:simpled}. The perturbation expansion of the
logarithmic term in~(\ref{eq:D}) contains diagrams constructed of specific
single skeleton blocks $\Sigma_i({\bf q})$ supplied by factors
$\zeta^{n_i}$. At $\zeta=1$ this expansion is identical to that on
both sides of Eq.~(\ref{eq:formal}). Now, any diagram is supplied with an
additional factor which is the product of factors of kind
$\zeta^{n_i}$ coming from all specific single skeleton blocks
of this diagram, yielding the resulting additional factor
$\zeta^{m_i-1}$ (consistent with the definition of $m_i$) or, according
to~(\ref{eq:formal}), the weight $m_i \, \zeta^{m_i-1}$ of the $i$--th skeleton
diagram. On the other hand, the same diagram appears in the perturbation
expansion $\sum\limits_i^{\hspace*{2ex} \prime} C_i \, \zeta^{m_i-1}$ of
$D^*(G,\zeta)$ with the weight $\zeta^{m_i-1}$ and in the perturbation
expansion of $\zeta \frac{\displaystyle \partial}
{\displaystyle \partial \zeta} D^*(G,\zeta)$ with the
weight $\left( m_i-1 \right) \, \zeta^{m_i-1}$. This, obviously, leads to
the identity of perturbation expansions in the left--hand side and in the
right--hand side of~(\ref{eq:D}), which proves the statement.

 We have defined $D^*(G)$  by
\begin{equation} \label{eq:DD}
D^*(G) = D^*(G,1) \; ,
\end{equation}
which, obviously, provides the perturbation expansion of $D^*(G)$ defined before.
Based on Eqs.~(\ref{eq:D}) to~(\ref{eq:DD}), we can calculate
$D^*(G)$ without explicit counting of the linear chains.

At the final step of our diagrammatic transformations, we make
the summation over the linear chains contained in the single
skeleton block following the method in Sect.~\ref{sec:exG},
with the only difference that blocks
{\mbox \ssb,} instead of {\mbox \selfe,} are considered.

As a result, the new perturbation expansion of $\Sigma({\bf q},\zeta)$ is
\begin{eqnarray}
&& \Sigma({\bf q},\zeta) = \dg + \zeta \ddg
+ \zeta^2 \left\{ \dddg  \right. \label{eq:ee} \\ \vspace*{2ex}
&& \left. + \dddgg + \dddggg + \dddgggg \right\} + ... \;, \nonumber
\end{eqnarray}
where the perturbation expansion of the waved line represents the sum over
the linear chains of all possible lengths $N$, i.~e., the sum of the
geometrical progression,
\begin{equation} \label{eq:wl}
\wlq \stackrel{\mbox{def}} = -u_{\bf q} V^{-1} / [1-2 \Sigma({\bf q},\zeta)]
= -u_{\bf q} V^{-1} \sum\limits_{N \ge 0} [2 \Sigma({\bf q},\zeta)]^N \; .
\end{equation}
Acording to this procedure, the new diagrams do not contain parts
like \selfe and/or \mbox{\wblock,} i.~e., they are single blocks and
skeleton diagrams with respect to both solid and waved lines.
By this the grouping of diagrams is completed.
Such a grouping of diagrams is unique
in the sense that it allows to analyze all diagrams simultaneously, considering
asymptotic equations related to critical phenomena.

    It should be noted that the actual grouping of skeleton diagrams
cannot be extended trivially to include even the
simplest $\varphi^6$ term. If we include, e.~g.,
the $\varphi^6$ vertex {\mbox \sixv,} subblocks of the kind \blmix are
possible ( {\mbox \blo,} in particular), therefore, the cyclical coupling
of single skeleton blocks \ssb (see Sect.~\ref{sec:rsd}) yield diagrams
which are not skeleton diagrams anymore.
For example, the
diagram \bllo represents one of the diagrams involved in {\mbox \ssb.} The
cyclical coupling of two such blocks is not a skeleton diagram because it
can be split in two blocks \bloo with two outer solid lines.

\subsection{Equations for calculation of the correlation function}
\label{sec:Dyson}

 In this section equations are considered from
which $\Sigma({\bf q},\zeta)$  and $G({\bf k})$ can, in principle, be
calculated. Although this quantity is not defined unambiguously by the
perturbation expansion, we can use any of possible functions having the
expansion~(\ref{eq:ee}). Only function $\vartheta({\bf k})$ in
Eq.~(\ref{eq:sig}) can be finally affected by the specific choice. We have
defined $\Sigma({\bf q},\zeta)$  by equations
\begin{eqnarray}
\Sigma({\bf q},\zeta)&=&\Sigma^{(0)}({\bf q}) +
\int\limits_0^{u^{-p}} e^{-t_1} dt_1
\int\limits_0^{u^{-p}} e^{-t_2} B({\bf q},\zeta,t_1 t_2) \,dt_2 \;,
\label{eq:one}\\
B({\bf q},\zeta,t)&=&\sum\limits_{m=1}^{\infty}
\frac{\zeta^m t^m}{(m !)^2} \, \Sigma^{(m)}({\bf q},\zeta) \; ,
\label{eq:two}
\end{eqnarray}
where $\Sigma^{(m)}({\bf q},\zeta)$ represents the sum of diagrams of the
$m$--th order ($m \ge 0$) in (\ref{eq:ee}), and $p$ is a constant having the
value $0<p<1/2$. Term $\Sigma^{(0)}({\bf q})$ in~(\ref{eq:one}) is separated to ensure
that the boundary condition $\Sigma({\bf q},0)=\Sigma^{(0)}({\bf q})$ is
satisfied when the first diagram is retained in~(\ref{eq:ee}).
Here and in our further considerations expansion in the vicinity of
the point $u=+0$ is used.
Equations~(\ref{eq:one}) and (\ref{eq:two}) yield function
$\Sigma({\bf q},\zeta)$ which has the perturbation expansion equivalent
(i.~e., the same, if represented in terms of $u$) to~(\ref{eq:ee}), and to
$\Sigma({\bf q},\zeta)=\sum\limits_i \Sigma_i({\bf q}) \, \zeta^{n_i}$,
since $\int\limits_0^{u^{-p}} e^{-t} t^m dt = m!
+o \left( u^{-pm} \exp \left(-u^{-p} \right) \right)$ and
$B({\bf q},\zeta,t) \sim \zeta u t$  (for $0<t \le u^{-2p}$) hold at
$u \to 0$ and $0<p<1/2$. The latter ensures convergence of
integrals~(\ref{eq:one}) at $u \to 0$. This relation holds because the actual
expansion parameter in~(\ref{eq:two}) at $u \to 0$ is $\zeta u t$ and this
sum converges absolutely, as discussed below. At large $m$ the number of
terms of the $m$--th order increases
approximately as $m^{\gamma} 6^m m!$ with some constant $\gamma$ and these
terms behave as $-(-b)^m$ where $b>0$.
According to this consideration terms in~(\ref{eq:two}) can be approximated as
$const \cdot m^{\gamma} (-b \zeta t)^m /m!$ with new (6 times larger) value of
$b$. Consequently, the perturbation sum of $\Sigma({\bf q},\zeta)$
diverges at any values of parameters, whereas~(\ref{eq:two}) converges
absolutely at any given values of $u$, $\zeta$ and $t$. Therefore, we can find
$B({\bf q},\zeta,t)$ from~(\ref{eq:two}) with any desired
accuracy (the larger are $\zeta$ and $t$, the larger is the number of terms
to be counted). Then we can find $\Sigma({\bf q},\zeta)$
from~(\ref{eq:one}).

The Dyson equation for
$G({\bf k})$ following from Eqs.~(\ref{eq:Sigdef}) and (\ref{eq:sig}) is
\begin{equation} \label{eq:Dyson}
\frac{1}{2 G({\bf k})} = r_0 + ck^2
- \frac{\partial D(G)}{\partial G_i({\bf k})} + \vartheta({\bf k}) \; .
\end{equation}
The same equation with term $\vartheta({\bf k})$ neglected has been obtained
in Ref.~\cite{MK1}.

  Our further analysis is limited to the case $u({\bf x})=u \delta({\bf x})$
or $u_{\bf k}=u$, where $u>0$. In this case,
from~(\ref{eq:Dyson}), (\ref{eq:Dg}), and (\ref{eq:DD}) we obtain
\begin{equation} \label{eq:Gk}
1/(2G({\bf k})) = r_0+ck^2 +2u \widetilde G + R({\bf k})+
\vartheta({\bf k}) \; ,
\end{equation}
where
\begin{equation}
\widetilde G = \left<\varphi^2({\bf x}) \right>
= V^{-1} n \sum\limits_{\bf k} G({\bf k}) \; ,
\end{equation}
\begin{equation} \label{eq:Rk}
R({\bf k})=- \frac{\partial D^*(G,1)}{\partial G_i({\bf k}) } \; .
\end{equation}
All terms in Eq.~(\ref{eq:Gk}) are well defined. $D^*(G,\zeta)$ and,
consequently, $R({\bf k})$ is defined by Eqs.~(\ref{eq:D}) to (\ref{eq:two}).
According to the definition, $\vartheta({\bf k})$ is a quantity
which have to be included to obtain an exact equation which is satisfied by the exact
correlation function $G({\bf k})$ given by the statistical integrals.
We know that $\vartheta({\bf k})$ does not contribute to the formal expansion
of $G({\bf k})$ in  $u$  power series.
It means that $\lim\limits_{u \to 0} \vartheta({\bf k},u) u^{-\tau} =0$
holds for any positive $\tau$.
Our equations have an obvious physical solution
$\Sigma({\bf k},\zeta) \simeq \Sigma^{(0)}({\bf k})$,
$D(G) \simeq D^{(0)}(G)$, and $G({\bf k}) \simeq G_0({\bf k})$
at $r_0>0$ and $u \to 0$, which agree with the true (exact)
$G({\bf k})$. Analytic continuation to arbitrary $r_0$ value is
possible if one starts with a finite volume $V$ and consider
the thermodynamic limit $V \to \infty$ afterwards.

\section{Asymptotic solution and critical exponents}
\label{sec:myexp}
\subsection{Determining the critical exponents at $u \to 0$}
\label{sec:detex}

We will show in this section how the true asymptotic solution at
$k \to 0$ (at the critical point $T=T_c$) can be found from the simplified
equations where $u$ tends to zero.
The right--hand side of~(\ref{eq:Gk}) vanishes at ${\bf k}={\bf 0}$
and $V \to \infty$ at the critical point and, thus, we have
\begin{equation} \label{eq:Gkcr}
1/(2G({\bf k})) = ck^2 + R({\bf k}) - R({\bf 0})
+\vartheta({\bf k}) -\vartheta({\bf 0}) \hspace{3ex}
\mbox{at} \hspace{2ex} T=T_c \; .
\end{equation}
The asymptotic of the correlation function at $k \to 0$ in this case is
$G({\bf k}) \simeq a \, k^{-\lambda}$ where $a$ is a constant
and $\lambda=2-\eta$, $\eta$ being the critical
exponent. The case of the spatial dimensionality $d<4$ is considered.
The correlation function is well described by
$G({\bf k}) = a \, k^{-\lambda}$ within some
critical region $k<k_{crit}(u)$, where $k_{crit}(u)$ tends to zero at
$u \to 0$, since at $u=0$ the Gaussian approximation
with $\lambda=2$  is the solution of~(\ref{eq:Gkcr}) for any $k$. Let us define the
effective value of $\eta$  at some $k= \tilde k$ by
\begin{equation}
\left. \eta \left( \tilde k, u \right)=-\partial \left( \ln \left[
k^{-2} G^{-1}({\bf k}) \right] \right) / \partial(\ln k)
\right|_{k=\tilde k} \; .
\end{equation}
 Then, $k_{crit}(u)$ is defined by
\begin{equation} \label{eq:kcr}
\left| \eta \left( k_{crit}(u),u \right) -\eta \right| =
\varepsilon \; ,
\end{equation}
where $\varepsilon$ is a sufficiently small constant. According to the
universality hypothesis, we have $\lim\limits_{k \to 0} \eta(k,u)= \eta$
for any positive $u$. We have also
$\lim\limits_{u \to 0} \eta \left( u^r k_{crit}(u),u \right)= \eta$
at $r>0$, since the critical exponent is determined at $k \ll k_{crit}(u)$
if $u \to 0$, and, therefore, corresponds to the asymptotic solution at
$k \to 0$. Based on a non--perturbative analysis provided in
Appendix~B, we have shown that
\begin{equation} \label{eq:llim}
\lim\limits_{u \to 0} \left( u^s / k_{crit}(u) \right) = 0 \; ,
\end{equation}
holds at large enough $s$.
Quantity $k_{crit}(u)$  can be related to the region where the correlation
function is well approximated by the asymptotic expansion
$G({\bf k})= \sum\limits_l b_l \, k^{-\lambda_l}$  of any
given number of terms. In this case $k_{crit}(u)$  is defined by the
condition that the approximation error at $k=k_{crit}(u)$  corresponds to the
variation of the smallest $\lambda_l$ by some small $\varepsilon$.
Following the method in Appendix~B, the statement~(\ref{eq:llim}) holds in
this case too,
and coefficients $c_l(u)$ in the asymptotic expansion
$1/G({\bf k})=\sum\limits_l c_l(u) \, k^{2 \lambda-\lambda_l}$
(where $l \ge 0$ and $\lambda_0 \equiv \lambda$)  meet the condition
$\lim\limits_{u \to 0} \left( u^{s_l} / c_l(u) \right) = 0$
at large enough $s_l$.

  According to the discussed property of $c_l$, the following is true.
If we assume that, at $k=u^r k_{crit}(u)$ and $u \to 0$,
$\omega(k,u)=\vartheta({\bf k})-\vartheta({\bf 0})$ is either compatible with
some term $c_l \, k^{2 \lambda-\lambda_l}$, or is much larger than all these
terms, then $\lim\limits_{u \to 0} \left[ u^{\tau} / \omega
\left(u^r k_{crit}(u),u \right) \right]=0$ holds for large enough values
of $\tau$. On the other hand, the basic property of $\omega(k,u)$ is that
$\lim\limits_{u \to 0} \left( u^{\tau} / \mid \omega(k,u) \mid \right) = \infty$
holds for any $\tau$ at any arbitrarily small, but fixed  $k$
(see the end of Sect.~\ref{sec:Dyson}). From this and
Eq.~(\ref{eq:llim}) we
conclude the following: if the above assumption is true, then, at
$u \to 0$, $\mid \omega(k,u) \mid$
decreases faster than any negative power of $k$ if $k$ is increased within
some region of infinitely small, but larger than $u^r k_{crit}(u)$ values
of $k$. On the other
hand, the only essential singularity of the correlation function is at
${\bf k}={\bf 0}$, which means that the above assumption leads to unphysical
conclusions regarding behavior of $G({\bf k})$, calculated from
Eq.~(\ref{eq:Gkcr}), unless the sharp decrease of $\mid \omega(k,u) \mid$
is compensated by the corresponding variation in $R({\bf k})-R({\bf 0})$.
However, our further analysis strongly supports the idea that
$R({\bf k})-R({\bf 0})$ is a well defined smooth function of $k$
which behaves like some power of $k$ at $k \to 0$.
Thus, the compensation is impossible and the
discussed here assumption is false, i.~e., the opposite is true:
at $k=u^r k_{crit}(u)$ and $u \to 0$ the term $\omega(k,u)$
is negligible compared to any of corrections to scaling in the asymptotic
expansion of $1/G({\bf k})$.

In principle, critical
exponents can be found by calculating $G({\bf k})$
within $k \in [u^r k_{crit}(u);\Lambda]$  directly from
Eq.~(\ref{eq:Gkcr}) where the term $\vartheta({\bf k})-\vartheta({\bf 0})$
is neglected, followed by extrapolation of the
results to smaller values of $k$ in the form
$G({\bf k})=\sum\limits_l b_l \, k^{-\lambda_l}$. At $u \to 0$ this yields
true (exact) correlation function, since Eq.~(\ref{eq:Gkcr}) is exact
(according to the definition),
term $\vartheta({\bf k})-\vartheta({\bf 0})$  is negligible, and the lower
marginal value of the considered interval is infinitesimal
compared to $k_{crit}(u)$.  According to the above analysis, this method
yields exact critical exponents $\lambda_l$.

\subsection{Scaling properties of the main terms at $T=T_c$}
\label{sec:Gmain}

It is impossible to calculate precisely all terms in
Eq.~(\ref{eq:two}). However, this is an unique feature of
our reorganized diagram expansion that all terms have common scaling
properties where the order of diagram does not appear as a
relevant parameter. Our claims are based on the proof of these
scaling properties at $d=2, 3$ for separate terms and also for the whole
sum~(\ref{eq:two}) not cutting the series.
To simplify the notation, $n$
is considered as a fixed parameter not included in
the list of arguments.

We introduce a lower limit of the wave vector $k_{min}$ to simulate a
finite--size effect (as if the linear size of the system
would be $L= 2\pi / k_{min}$) in calculation of $R({\bf k})$,
which ensures the convergence of any ${\bf k}$--space integral.
Finally, we consider the limit $k_{min} \to 0$ in the equation for
$R({\bf k})$.

   Since $\eta >0$ holds for $d<4$, the term $ck^2$ in~(\ref{eq:Gkcr})
gives a small correction which is by factor $k^{\eta}$ smaller than the main
term $R({\bf k})-R({\bf 0})$. Term "1" in~(\ref{eq:D}), (\ref{eq:Db}), and
(\ref{eq:wl}) causes a small correction to $R({\bf k})$ as well.
This can be checked easily by a direct calculation as in
Ref.~\cite{MK2}, if only the first term in~(\ref{eq:ee}) is retained.
We prove here that a selfconsistent solution of our equations can be found
where this condition is satisfied with account for all terms.

  First, let us consider the dominant behavior of
$\Sigma({\bf q},\zeta)$ calculated at
$G({\bf k})= a \, k^{-\lambda}$. A property of the asymptotic solution
is such that
\begin{equation} \label{eq:Sigmain}
\Sigma({\bf q},\zeta)= a^2 \times q^{d-2 \lambda} \,
\Psi \left( \Lambda/q,k_{min}/q,\zeta,\lambda,d,u \right)
\end{equation}
holds, where $\Psi$ is a function of the given arguments only.
Such a solution is
possible since each term in~(\ref{eq:two}) has the scaled form
like~(\ref{eq:Sigmain}) if~(\ref{eq:wl}) is substituted
by~(\ref{eq:Sigmain}) neglecting "1". It becomes obvious after
the following manipulations: all the sums are replaced by integrals
(the standard procedure),
$V^{-1} \sum_{\bf k} \Longrightarrow (2 \pi)^{-d} \int d^d k$,
and all the wave vectors in~(\ref{eq:ee}) or (\ref{eq:two}) are normalized
to the current value of $q$.
This produces factor $a \, q^{-\lambda}$ corresponding to each of the solid
lines, factor $a^{-2} q^{2 \lambda -d}$ corresponding to each of the waved
lines, and factor $q^d$  corresponding to each
integration for any diagram of~(\ref{eq:ee}). As a consequence, the
resulting factor is $a^2 q^{d-2 \lambda}$ irrespective of the order of
diagram $m$ (there are $2m+2$ solid lines, $m$ waved lines, and
$m+1$ integrations). Thus, $a^2 q^{d-2\lambda}$ appears
as a common prefactor for the whole sum~(\ref{eq:two}), and the problem
reduces to summation of scaling functions depending merely on
$\Lambda/q$, $k_{min}/q$, $\zeta$, $\lambda$, $d$, and $u$,
which leads to~(\ref{eq:Sigmain}).
In this case arguments $\Lambda/q$ and $k_{min}/q$ represent the upper
and the lower limits of integration for normalized wave vectors related to
the solid lines.

Formally we could allow other kind of solutions, but the true
result must coincide with the
asymptotic expansion $\Sigma({\bf q},\zeta)=
\sum_m \zeta^m \, \Sigma^{(m)} ({\bf q})$ at $\zeta \to 0$
(consistent with the boundary condition
$\Sigma({\bf q},0)=\Sigma^{(0)}({\bf q})$), which shows
that~(\ref{eq:Sigmain}) is the only possibility.
If the correction term "1" in Eq.~(\ref{eq:wl}) is omitted,
then any of the expansion coefficients has the scaled form with
factor $a^2 q^{d-2 \lambda}$ multiplied by some function
of $\Lambda/q$, $k_{min}/q$, $\lambda$, $d$, and $u$.
 This is proved by induction over $m$: it
holds at $m=0$; if it holds for terms up to the $m$--th order, then it holds
for terms of the $(m+1)$--th order, calculated from diagrams with no more
than $m+1$ waved lines (expanded in terms of $\zeta$).
Thus, the asymptotic solution at $\zeta \to 0$ is unambiguous and
has the scaled form~(\ref{eq:Sigmain}), which is a property of
the solution as well at finite $\zeta$.

It is purposeful to seek a solution for
$\partial \Sigma({\bf q},\zeta) / \partial G_i({\bf k})$ in form\linebreak
$V^{-1} a k^{-\lambda} \,
Y \left( {\bf q}/k,{\bf k}/k,\Lambda/k,k_{min}/k,\zeta,\lambda,d,u \right)$
using Eqs.~(\ref{eq:one}) and (\ref{eq:two}), where
$\Sigma({\bf q},\zeta)$ is considered as a quantity which is already known.
The latter can be represented in the form
$a^2 k^{d-2 \lambda} \, \widetilde
\Psi \left( q/k,\Lambda/k,k_{min}/k,\zeta,\lambda,d,u \right)$
obtained from~(\ref{eq:Sigmain}) by changing variables
$q \to k \cdot \left( q/k \right)$. This yields a selfconsistent equation
for the unknown function $Y$. The arguments of this function only are
contained therein, since all terms
$\partial \Sigma^{(m)}({\bf q},\zeta) / \partial G_i({\bf k})$ in the sum
for $\partial B({\bf q},\zeta,t) / \partial G_i({\bf k})$, obtained
from~(\ref{eq:two}), have the form $V^{-1} a k^{-\lambda}$ multiplied by
some function of these arguments. In a similar way, we can find from the
closed equations~(\ref{eq:D}) and (\ref{eq:Db}) a selfconsistent solution
$\partial D^*(G,\zeta)/\partial G_i({\bf k})= a^{-1} k^{\lambda} \,
\bar \phi \left( \Lambda/k,k_{min}/k,\zeta,\lambda,d,u \right)$.
The analysis of the asymptotic solution at $\zeta \to 0$ shows that
discussed above are the true solutions meeting~(\ref{eq:Db}).

  Consider now the limit $k_{min} \to 0$. The nonzero lower limit of
integration has been introduced to avoid the divergence of ${\bf k}$--space
integrals in~(\ref{eq:ee}) considering the formal
expansion in terms of $\zeta$. If the selfconsistent solution of
Eqs.~(\ref{eq:one}) and (\ref{eq:two}) is
considered, then convergence of integrals at $k_{min}=0$ is ensured
since these equations provide the solution
of physical problem where the existence of thermodynamic limit
($k_{min} \to 0$) is doubtless, and this conclusion
agree with the formal analysis of our equations at $k_{min}=0$:
$\Sigma({\bf q},\zeta)$ diverges at $q \to 0$, but this term appears in the
denominator of~(\ref{eq:wl}), which ensures the convergence of
${\bf k}$--space integrals in~(\ref{eq:ee}). A selfconsistent solution with
exponentially diverging $\Sigma({\bf q},\zeta)$ at $q \to 0$ is
not possible, which ensures the convergence in~(\ref{eq:D}). Thus, we can set
$k_{min}=0$, which at $\zeta=1$ leads directly to the
relation
$a k^{-\lambda} \, R({\bf k})= \phi^* \left( \Lambda/k,\lambda,d,u \right)$,
where $\phi^*$ is a function exclusively of the given arguments.
Since $R({\bf 0})$ is a constant, the function $\phi^*$ can be represented as
$\phi^* \left( \Lambda/k,\lambda,d,u \right)= a \Lambda^{-\lambda}
R({\bf 0}) \cdot \left( \Lambda/k \right)^{\lambda} +
1/ \left[ 2 \phi \left( \Lambda/k,\lambda,d,u \right) \right]$.
We get
\begin{equation} \label{eq:Gmain}
G({\bf k})=a \, k^{-\lambda} \, \phi \left( \Lambda/k,\lambda,d,u \right) \;
\end{equation}
by substituting this into~(\ref{eq:Gkcr}).

   According to the consideration in Sect.~\ref{sec:detex}, the true value of
$\eta$ can be found from our equations in the limit $u \to 0$. In this case
behavior of $G({\bf k})$ within some region $k \sim u^r k_{crit}(u)$
and extrapolation to smaller values of  $k$ have to be considered. It
follows from the definition of $k_{crit}(u)$  that a stable solution in the form
$G({\bf k}) \simeq a k^{-\lambda}$ does exist at $k \sim u^r k_{crit}(u)$.
It means that $\phi$ does not depend on $\Lambda/k$ (or $k$) within
this region at $u \to 0$. Thus, the value of argument $\Lambda/k$
in~(\ref{eq:Gmain}) can be replaced by $\Lambda \, u^{-r} k_{crit}^{-1}(u)$.
On the other hand, universal positive value of $\eta$ is obtained
from Eq.~(\ref{eq:Gkcr}) at $u \to 0$, which means that a finite limit
$\lim\limits_{u \to 0} \phi \left( \Lambda \, u^{-r}
k_{crit}^{-1}(u),\lambda,d,u \right) = b(\lambda,d)$ exists.
So, we have an asymptotic equation
\begin{equation}
G({\bf k})=a \, k^{-\lambda} = a \, k^{-\lambda} \times b(\lambda,d)
\end{equation}
from which, in principle, the universal value of $\lambda$
can be found ( i.~e., $b(\lambda,d)=1$).

\subsection{Scaling properties at $T=T_c$ including correction terms}

  The analysis of the previous section can be extended by including
corrections to scaling. As it was mentioned, correction
$\varepsilon(k) \sim k^{\eta}$ is introduced by $ck^2$
in~(\ref{eq:Gkcr}). As regards term "1" in~(\ref{eq:D}),
(\ref{eq:Db}), and (\ref{eq:wl}), it yields (performing the same analysis
as for the main terms) a correction $\delta(k) \sim k^{2 \lambda -d}$
in~(\ref{eq:Gkcr}) corresponding to the contribution of the correction term
(with the factor $q^{2 \lambda -d})$ in equation
\begin{equation}
1-2 \Sigma({\bf q},\zeta) \simeq -2 \Sigma({\bf q},\zeta) \times
\left( 1- q^{2 \lambda -d} / \left[ 2a^2 \, \Psi \left(
\Lambda/q,k_{min}/q,\zeta,\lambda,d,u \right) \right] \right) \; .
\end{equation}
Here, and in equations~(\ref{eq:ee}) to
(\ref{eq:two}) $q$, $\Lambda/q$, and $k_{min}/q$ are considered as independent
variables, considering the limit $q \to 0$ at any given $\Lambda/q$,
and $k_{min}/q$. Similarly, in equations for $R({\bf k})$ and
$G({\bf k})$ $k$, $\Lambda/k$, and $k_{min}/k$ are considered as independent
variables. Finally, we set $k_{min}/k \to 0$ to fit the
thermodynamic limit, and then we consider the limit $u \to 0$ with the
simultaneous tending of $\Lambda/k$ to infinity (as in
Sect.~\ref{sec:Gmain}) to get the
asymptotic solution at a fixed $\Lambda$. The solution can
be expanded in terms of $\varepsilon(k)$ and $\delta(k)$ to yield an
asymptotic expansion at $k \to 0$
\begin{equation} \label{eq:Gasex}
G({\bf k})= \sum\limits_l b_l \, k^{-\lambda_l}
\end{equation}
where
\begin{equation} \label{eq:expc}
\lambda_l = \lambda - n_l \cdot \eta - m_l \cdot (2 \lambda -d)
\end{equation}
with $n_l$, $m_l$  = 0, 1, 2, ...  corresponding to the correction of order
$\varepsilon^{n_l} \delta^{m_l}$, $b_l$  being
expansion coefficients.

If the right hand side of
Eq.~(\ref{eq:Rk}) is substituted by~(\ref{eq:Gasex}), we get
\begin{equation}
R({\bf k})= \sum\limits_l \phi_l^* \left( \Lambda/k,\lambda,d,u
\right) \, k^{2 \lambda -\lambda_l}
\end{equation}
where $\phi_l^* \left( \Lambda/k,\lambda,d,u \right)$
are functions of the given arguments. They depend also on the
set of coefficients $b_l$. This is obtained by finding selfconsistent
solutions similarly as in
Sect.~\ref{sec:Gmain}, with the only difference that any given
number of corrections to the scaling is included.

    Since $R({\bf 0})$ is constant, $\phi_l^*$
behaves as $const \cdot \left( \Lambda/k \right)^{2 \lambda -\lambda_l}$ at
$k \to 0$, and herefrom it follows that $R({\bf 0})$ can be represented as
$R({\bf 0})= \sum\limits_l R_l \, \Lambda^{2 \lambda -\lambda_l}$ with
$R_l$ dependenging on $\lambda$, $d$, $u$, and coefficients $b_l$.
Based on the same logic as in Sect.~\ref{sec:Gmain}, functions
$\hat \phi_l \left( \Lambda/k,\lambda,d,u \right)
= \phi_l^* \left( \Lambda/k,\lambda,d,u \right) - R_l \,
\left(\Lambda/k \right)^{2 \lambda -\lambda_l}$ may be replaced by
$\hat \phi_l \left( \Lambda u^{-r} k_{crit}^{-1}(u),\lambda,d,u \right)$
at $u \to 0$ to obtain the asymptotic solution~(\ref{eq:Gasex}) at a fixed
$\Lambda$. This replacement is justified by the following argument.
If the amplitude $a$ ($\equiv b_0$) is considered as known (fixed)
quantity and $k \sim u^r k_{crit}(u) \to 0$, then
$\hat \phi_l \left( \Lambda/k,\lambda,d,u \right)$  tends to some function
$\widetilde \phi_l (\lambda,d,u)$ and
$\hat \phi_l \left( \Lambda/k,\lambda,d,u \right)-\widetilde \phi_l (\lambda,d,u)$
tends to zero faster than any positive power of $k$, since only in this case
exponents in the right--hand side of~(\ref{eq:Gkcr}) are not affected by
short--wave fluctuations and are the same as those in the left--hand side of
the equation.

To show this precisely, we prove the following statement:
if in the considered limit
$\hat \phi_l \left( \Lambda/k,\lambda,d,u \right)$ tends to
$\widetilde \phi_l (\lambda,d,u)$, then the tending is faster than
$\left( \Lambda/k \right)^{-\sigma}$, where $\sigma$ is any finite and positive
constant. Really, if $G({\bf k})$ in the right hand side of
Eq.~(\ref{eq:Gkcr}) is replaced by $G({\bf k})- \delta G({\bf k})$, where
$\delta G({\bf k})=G({\bf k}) \cdot \left( k/ \Lambda' \right)^m
/ \left[ 1+ \left( k/ \Lambda' \right)^m \right]$, $\Lambda' < \Lambda$,
and $m \to \infty$, then this is equivalent
to the shift of the upper limit of wave vector magnitude from
$\Lambda$ to $\Lambda'$ at a fixed amplitude $a$. On the other hand, the quantity
$\delta G({\bf k})$ can be treated as any other
correction term. If for arbitrary ($l$--th) correction $\hat \phi_l$ tends
to $\widetilde \phi_l$, then at $k \sim u^r k_{crit}(u) \to 0$ the quantity
$\delta G({\bf k})$ produces in the right hand side of
Eq.~(\ref{eq:Gkcr}) a correction term of order $k^{\lambda+m}$ where
$m \to \infty$. Thus, the shift of $\Lambda$  produces a
correction smaller than $k^{\sigma}$ at any finite $\sigma$,
which proves the statement.

   According to physical arguments, the condition that $\hat \phi_l$ tends
to $\widetilde \phi_l$, obviously, is satisfied. It means that the main
contribution to $\hat \phi_l \left( \Lambda/k,\lambda,d,u \right)$ in the
equation for $1/G({\bf k})$  at $k \to 0$ is provided by the integration
over small wave vectors (therefore the result is almost independent on
$\Lambda$), i.~e., the critical behavior is governed by the long--wave
fluctuations.
Besides, this condition means that the asymptotic solution in approximation
$G({\bf k})= \sum_{l=0}^{m} b_l \, k^{-\lambda_l}$,
including $m$ correction--to--scaling terms ($m$=0, 1, 2, etc.), is stable
with respect to (i.~e., is not changed by) higher order
corrections, which is reasonable and expected.

   As regards corrections, we allow a possibility that any of corresponding
expansion coefficients can be zero, our analysis is correct in this case.

\subsection{Asymptotic solution at $T \to T_c$}

In this section we have extended our scaling analysis to describe
the critical behavior of the model when approaching the critical
point from higher temperatures, i.~e., at positive $\Delta=T-T_c \to 0$.

The critical
exponents cannot be affected by short--wave fluctuations. According to this
idea, the contribution of sufficiently large $k$ may be neglected in
equations for $1/G({\bf 0})$ and $[1/G({\bf k})]-[1/G({\bf 0})]$
\begin{equation} \label{eq:A0}
\frac{1}{2G({\bf 0})} = \frac{dr_0}{dT} \cdot \Delta + \frac{2u}{V}
\sum\limits_{\bf k}
\left[ G({\bf k}) - G^*({\bf k}) \right]
+ R({\bf 0}) - R^*({\bf 0}) \; ,
\end{equation}
\begin{equation} \label{eq:Ak}
\frac{1}{2G({\bf k})} - \frac{1}{2G({\bf 0})}
= R({\bf k}) - R({\bf 0}) + ck^2
\end{equation}
obtained from Eq.~(\ref{eq:Gk}) omitting the irrelevant correction
$\vartheta({\bf k})$ and assuming that
$r_0(T)=r_0(T_c)+ (dr_0 / dT) \cdot \Delta$ is the only parameter
in~(\ref{eq:Hp}) which depends on temperature.
In Eq.~(\ref{eq:A0}), $G^*({\bf k})$ is the value of
$G({\bf k})$ at $T=T_c$ and $R^*({\bf 0})$ is the value of $R({\bf 0})$
calculated at $G({\bf k}) = G^*({\bf k})$. This equation represents the
condition that $1/G({\bf 0})$ vanishes at $T=T_c$.

   Considering the solution at $T=T_c$, we have concluded that terms
in the right--hand side of equation for $1/(2G({\bf k}))$, calculated at a
fixed $G({\bf k})$ (i.~e., at fixed amplitudes $b_l$ in~(\ref{eq:Gasex})),
are not sensitive to a variation in the upper
limit of the wave vector magnitude $\Lambda$ at $u \to 0$ if
$\Lambda/k \sim u^{-r} k_{crit}^{-1}(u)$ holds, where  $r$ is any
positive constant.
In this case unambiguous solution insensitive to the short--wave
fluctuations is obtained, based on the asymptotic
expansion~(\ref{eq:Gasex}), if the main amplitude $b_0$ is
considered as a known (fixed) quantity.
The asymptotic solutions at $T=T_c$ and
$T>T_c$ join at $k \sim 1/\xi$, where $\xi \sim \Delta^{-\nu}$ is
the correlation length, therefore, the solution of
Eqs.~(\ref{eq:A0}) and~(\ref{eq:Ak})
at $T \to T_c$ for $k \sim 1/\xi$ is insensitive to variation of
$\Lambda$ within some region $\Lambda \sim \xi^{-1} u^{-r} k_{crit}^{-1}(u)$
(at $u \to 0$), if calculations are performed at fixed $G^*({\bf k})$
(i.~e., at fixed $b_0$). It is supposed that $k_{crit}(u)$ is
determined at a fixed upper integration limit $\Lambda'$, and
$\Lambda$ is smaller than, but comparable with $\Lambda'$. The latter
condition is satisfied for values of $\Delta$ which are smaller
than, but comparable with $B \, u^{r/\nu} \Delta_{crit}(u)$, where $B$
is appropriate constant and
$\Delta_{crit}(u) \sim \left[ A(u) k_{crit}(u) \right]^{1/\nu}$ is the
width of the critical region inside of which the correlation function is
described by $G({\bf k})=\xi^{\lambda} g({\bf k} \xi)$ with the
relative error not exceeding some small given value. According to our
definition $\xi \simeq A(u) \Delta^{-\nu}$ holds, and the above relation for
$\Delta_{crit}(u)$ is true since the width of the
critical region for $\xi^{-1}$  is proportional to $k_{crit}(u)$  due to the
joining of asymptotic solutions at $k \sim 1/\xi$.
We can conclude from this discussion that the region $k>C(u)/\xi$
with $C(u)=u^{-r} k_{crit}^{-1}(u)$ corresponds to negligible
short--wave fluctuations if the solution inside the asymptotic region
$k \sim 1/\xi$ is considered at $u \to 0$ and
$\Delta \sim B \, u^{r/\nu} \Delta_{crit}(u)$.
In such a way, we may neglect
the short--wave fluctuations by formally setting
$\Lambda=\hat C(u) \Delta^{\nu}$
(i.~e., $G({\bf k})=0$ and $G^*({\bf k})=0$ at $k> \hat C(u) \Delta^{\nu}$)
in the right--hand side of Eqs.~(\ref{eq:A0}) and~(\ref{eq:Ak}) where
$G^*({\bf k})$ has been calculated before this procedure at the true
(constant) upper limit~$\Lambda'$, and $\hat C(u) = C(u)/A(u)$.
According to the above discussion, this method provides correct
correlation function at $k \sim 1/\xi$, which means that it yields true
critical exponents.

In this case the asymptotic solution can be found in the form
\begin{equation} \label{eq:as}
G({\bf k})= \sum\limits_{l \ge 0} \Delta^{-\gamma+\gamma_l} \,
g_l({\bf k} \Delta^{-\nu}) \;,
\end{equation}
where exponents $\gamma_l$ are related to those given
by~(\ref{eq:expc}) in the way predicted by the scaling
hypothesis, i.~e., $g_l(y)$ behave like
$g_l(y) \simeq b_l \, y^{-\lambda_l}$ at $y \to \infty$
to yield~(\ref{eq:Gasex}) at $\Delta \to 0$.
Thus, the exponents in~(\ref{eq:as}) are
\begin{eqnarray}
\gamma &=& \lambda \nu = (2- \eta) \, \nu \;, \label{eq:scre} \\
\gamma_l &=& n_l \, \delta_1 + m_l \, \delta_2 \;, \label{eq:gl}
\end{eqnarray}
where
\begin{eqnarray}
\delta_1 &=& 2 \nu -\gamma > 0 \;, \label{eq:del1} \\
\delta_2 &=& 2 \gamma- d \nu >0 \;. \label{eq:del2}
\end{eqnarray}
Here the main term is given by $l=0$ and $\gamma$ is the susceptibility
exponent. The $l$--th term with $l>0$ represents the correction
of order $\varepsilon^{n_l} \delta^{m_l}$, where
$\varepsilon(\Delta)=\Delta^{\delta_1}$ and
$\delta(\Delta)=\Delta^{\delta_2}$.

   This result is obtained by a scaling analysis similar to
that we have made at $T=T_c$. The only difference is that wave
vectors are normalized to $\Delta^{\nu}$, but not to the current
value of $q$ (in equation for $\Sigma({\bf q},\zeta)$) or $k$
(in equation for $G({\bf k})$). Besides, the lower limit of
integration may be set $k_{min}=0$ from the very beginning since
there is no singularity at ${\bf k}={\bf 0}$.
In such a way, retaining only the main term,
we prove the following scaled form for the relevant quantities
calculated at $\Lambda= \hat C(u) \Delta^{\nu}$:
$\Sigma({\bf k},\zeta)= \Delta^{-2 \gamma+ d \nu} \Psi({\bf k}',\zeta)$,
$\partial \Sigma({\bf q},\zeta) / \partial G_i({\bf k})
= \Delta^{-\gamma} \, V^{-1} Y({\bf q}',{\bf k}',\zeta)$,
$\partial D^*(G,\zeta) / \partial G_i({\bf k})
= \Delta^{\gamma} \, \widetilde r({\bf k}',\zeta)$, and
$R({\bf k})= \Delta^{\gamma} \, r({\bf k}')$,
where ${\bf k}'={\bf k} \Delta^{-\nu}$, and at any fixed $u$, $d$,
$n$, and $g_0({\bf k} \Delta^{-\nu})$ the scaling functions depend on
the given arguments only. Including corrections produced by
$ck^2$ in Eq.~(\ref{eq:Gk}) (correction of order
$\Delta^{\delta_1}$) and "1" in Eq.~(\ref{eq:wl})
(correction of order $\Delta^{\delta_2}$),
we obtain the asymptotic expansion in the scaled form
\begin{equation} \label{eq:R}
R({\bf k}) = \sum\limits_{l \ge 0} \Delta^{\gamma+\gamma_l}
\, r_l({\bf k}') \;
\end{equation}
with exponents $\gamma_l$ defined by Eq.~(\ref{eq:gl}).
If Eqs.~(\ref{eq:A0}) and~(\ref{eq:Ak}) are substituted
by~(\ref{eq:R}) at $\Lambda= \hat C(u) \Delta^{\nu}$,
then we obtain a solution where no other correction exponents
appear.

\subsection{Possible values of critical exponents}

By using the equations and scaling relations obtained in previous
sections, here we derive our central result -- the set of possible
values for critical exponents.

It is reasonable to assume that
$\gamma > 1$, which leads to the conclusion that $(dr_0/dT) \cdot \Delta$
in Eq.~(\ref{eq:A0}), where $\Lambda= \hat C(u) \Delta^{\nu}$,
is compensated by one of the terms coming from the asymptotic expansion
\begin{equation} \label{eq:tram}
2u \, V^{-1} \sum\limits_{k< \hat C(u) \Delta^{\nu}}
G({\bf k})- G^*({\bf k})
= \sum\limits_{l \ge 0} B_l \, \Delta^{\gamma -\delta_2+ \gamma_l} \;.
\end{equation}
It means that
\begin{equation} \label{eq:ggg}
\gamma_m = 1- \gamma + \delta_2
\end{equation}
holds at some $m \ge 0$. Condition $\gamma_m < \delta_2$ follows herefrom,
since $\gamma>1$. Thus, with account for~(\ref{eq:gl}) we have
$\gamma_m= m \, \delta_1$, and Eq.~(\ref{eq:ggg}) becomes
\begin{equation} \label{eq:rel}
m \, \delta_1 - \delta_2 + \gamma =1 \;.
\end{equation}
 We need one more relation to determine the values of critical
exponents. We obtain this relation from calculation of specific
heat $C_V$ assuming the well known hyperscaling hypothesis
\begin{equation} \label{eq:hypersc}
\alpha + d \nu =2 \; .
\end{equation}
The singular part of the specific heat
behaves like $\Delta^{-\alpha}$ and
it can be related to the singular part of $\widetilde G$,
i.~e., to $\widetilde G - \widetilde G^*$ (where $\widetilde G^*$
is the value of $\widetilde G$ at $T=T_c$), as follows
\begin{equation} \label{eq:Cvs}
C_V \propto \frac{\partial}{\partial \Delta}
\left[ V^{-1} \sum\limits_{k<\Lambda'}
\left( G({\bf k})-G^*({\bf k}) \right) \right] \; ,
\end{equation}
where $\Lambda'$ is constant.
The latter relation follows from thermodynamics, taking into
account that
\begin{equation}
\frac{\partial}{\partial r_0} \left( \frac{F}{T} \right)
= - \frac{\partial \ln Z}{\partial r_0}
= V \left< \varphi^2({\bf x}) \right> \equiv V \widetilde G
\end{equation}
holds where $F$ is the free energy and $Z$ is the statistical sum,
consistent with the definition of Hamiltonian~(\ref{eq:Hp}) where
$r_0$ is the only parameter depending on temperature and the dependence
is linear.

 According to the universality hypothesis, critical exponents
do not depend on the coupling constant $u$, therefore the
exponents obtained by our method at $u \to 0$ can be used in
calculation of the singular part of
$C_V$ at a finite $u$.
Coefficients $B_l$ in~(\ref{eq:tram}) can be changed, not
changing $G({\bf k})$ at $k \sim 1/\xi$, if the solution at
$\Lambda=\Lambda'=const$ instead of the formal solution at
$\Lambda= \hat C(u) \Delta^{\nu}$ is considered. For instance,
a variation in $B_l$ due to the change in $G({\bf k})$ at
$k \sim \hat C(u) \Delta^{\nu} \gg 1/\xi$ (at $u \to 0$)
can be compensated by a
contribution coming from $R({\bf 0})-R^*({\bf 0})$ due to the
integration over $\hat C(u) \Delta^{\nu} < k < \Lambda'$.
Thus, the true values of the amplitudes in the expansion of
$\widetilde G - \widetilde G^*$ are unknown, and we allow
all the possibilities.

Thus, consider a finite $u$. Our basic idea is that
the contribution to~(\ref{eq:Cvs}) provided by the summation
over $k> C/\xi$ or $k > C \Delta^{\nu}$
at $C \to \infty$ (we consider the limit $\Delta \to 0$ at
a given $C$, which then is tended to infinity) with the true
correlation functions $G({\bf k})$ and $G^*({\bf k})$
(calculated at $\Lambda=\Lambda'$) cannot change the critical
exponent $\alpha$,
because the opposite would mean a violation of scaling relations
for critical exponents. However, we allow that a logarithmic
correction can be caused by this contribution. Thus, we can
replace the summation limit $k< \Lambda'$ in~(\ref{eq:Cvs})
by $k< C \Delta^{\nu}$. The first non--vanishing (i.~e.,
having nonzero amplitude at $C = \infty$) singular term
in the resulting asymptotic expansion represents the leading
singularity of $C_V$. Formally, a constant contribution also is
defined as singular (with $\alpha=0$) in the case if there is a
jump of $C_V$ at $T=T_c$ from one constant value to another, or
if a refined analysis reveals logarithmic singularity.
Thus, according to~(\ref{eq:Gasex}),
(\ref{eq:as}), (\ref{eq:ggg}),
and~(\ref{eq:Cvs}), all possible values of $\alpha$ are given by
\begin{equation} \label{eq:alpha}
\alpha = \gamma_m - \gamma_i
\end{equation}
where $i \ge 0$ is integer. In this case
$g_i(y) - b_i \, y^{- \lambda_i}$ (cf. Eqs.~(\ref{eq:as})
and~(\ref{eq:Gasex})) tends to zero at $y \to \infty$
and the tending is faster than $y^{-\sigma}$ where $\sigma < d$,
since the opposite would mean that
the critical exponent $\alpha$ is changed due to the contribution
of $k> C \Delta^{\nu}$. In the marginal case when
$g_i(y) - b_i \, y^{- \lambda_i} \sim y^{-d}$ holds this contribution
yields a logarithmic correction, i.~e.,
$C_V \sim \Delta^{-\alpha} \, \ln \Delta$. In such a way
our theory provides an explanation of the known logarithmic
singularity of the specific heat at $n=1$ and $d=2$.

We have restricted
our analysis to $\gamma>1$ and $\alpha> 1- \gamma$, which is very
reasonable assumption in view of the known results.
From~(\ref{eq:ggg}) we
obtain $1- \gamma = \gamma_m - \delta_2$, which, in this case, yields
$\alpha = \gamma_m -\gamma_i > \gamma_m - \delta_2$ or $\gamma_i< \delta_2$.
Then, combining~(\ref{eq:rel}) with~(\ref{eq:alpha})
and~(\ref{eq:hypersc}), with account for
definitions~(\ref{eq:del1}), (\ref{eq:del2}), and (\ref{eq:gl}),
the following set of possible values for critical exponents is obtained
\begin{equation} \label{eq:result}
\gamma = \frac{d+2j+4m}{d(1+m+j)-2j} \;; \hspace*{3ex}
\nu = \frac{2(1+m)+j}{d(1+m+j)-2j} \; ,
\end{equation}
where $m$ may have a natural value starting with 1 and $j$ is integer
equal or larger than $-m$.
This result is obtained by proving all the relevant scaling
properties not cutting the perturbation
series~(\ref{eq:two}), based on equations which allow to find
the exact critical exponents (Sec.~\ref{sec:detex}).
Thus, according to these arguments, Eq.~(\ref{eq:result})
represents all possible values for the exact critical exponents at
$d=2$ and $d=3$ in the cases where the
second--order phase transition with spontaneous
long--range ordering takes place.

Our analysis is not valid at $d \ge 4$ since $\delta_1$ and $\delta_2$
(Eqs.~(\ref{eq:del1}) and~(\ref{eq:del2})) are positive,
i.~e., $\Delta^{\delta_1}$ and $\Delta^{\delta_2}$ are
small corrections, merely at $d<4$. Besides, the
analysis in Appendix~B is not true at $d=4$.
Solutions at natural $n$ (the dimensionality of the order
parameter) only have a meaning, since our method is
strongly based on the proof that in the relevant asymptotical
region at $u \to 0$ the solution of our equations
agree with the exact correlation function defined by
the statistical integrals.
Although the formal perturbation expansion exists
at arbitrary $n$, such a method of proof would be meaningless
at a not natural $n$, since the exact correlation function
is not defined in this case, and we cannot guarantee that a formal
solution at arbitrary $n$ has all the correct properties
(e.~g., existence of the secon--order phase transition
and scaling relation~(\ref{eq:hypersc}))
which have been assumed to derive Eq.~(\ref{eq:result}).
Our predictions do not refer also to the case $n=0$.
This case is exceptional in view of our analysis,
since the term $2 \Sigma({\bf q},\zeta)$ in the denominator
in Eq.~(\ref{eq:wl}) which appears as the main term at $n \ge 1$
vanish at $n=0$.

In general, different values of $j$ and $m$
can correspond to different (natural) $n$, i.~e., $j=j(n)$ and $m=m(n)$.
It is easy to verify that at $j=0$ and
$m=3$ Eq.~(\ref{eq:result}) reproduces the
known~\cite{Baxter} exact results in two dimensions.
The known exact exponents for the spherical model~\cite{Baxter}
($n=\infty$) are obtained at $j(n)/m(n) \to \infty$. Although the
derivations are true for $d<4$, Eq.~(\ref{eq:result}) provides
correct result $\nu=1/2$ and $\gamma=1$ also at $d=4$.
It is reasonable to consider $d$ as a continuous parameter. This
leads to the conclusion that $m=3$ and $j=0$ are the correct values
 for the case $n=1$ not only at $d=2$, but also at $d=3$. In the
latter case we have $\gamma=5/4$ and $\nu=2/3$. The nearest
values of $\gamma$ and $\nu$ provided by Eq.~(\ref{eq:result}),
e.~g., at $j=1$ and $m=3$ or at $j=1$ and $m=4$ are then the most
probable candidates for the case $n=2$.
It is interesting to note that our prediction for the singularity
of specific heat $\alpha=0$ for the Ising model
($n=1$) agree with that made by Tseskis~\cite{Tseskis}
based on a fractal model.

\section{Comparison with Monte--Carlo and experimental \\
results and discussion} \label{sec:compare}

It is commonly believed that all more or less correct Monte Carlo (MC)
simulations confirm the values of critical exponents obtained from the
perturbation expansions based on the renormalization group.
This is not true. We have found that
some kind of MC simulations at the critical point, namely,
 the MC simulations of fractal configurations of Ising model~\cite{IS}
and the MC simulations of the energy density~\cite{SM} for the $XY$
model in reality do
not confirm the results of the RG theory, but provide the values
of critical exponents which are very close to those we predicted.

\begin{figure}
\centerline{\psfig{figure=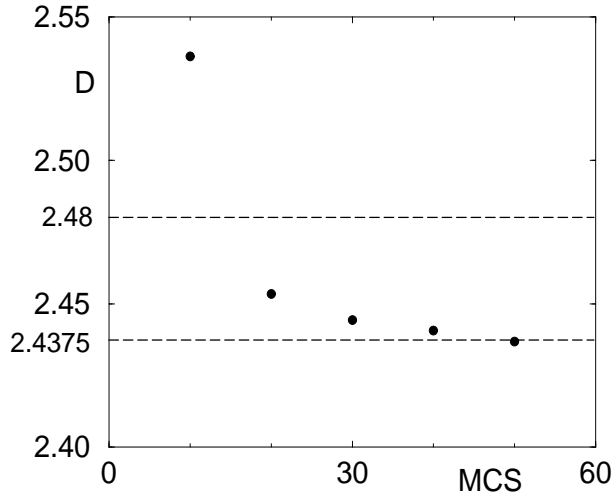,width=8cm,height=6.5cm,angle=-90}}
\caption{\small Fractal dimensionality $D$ of the three dimensional Ising
model at the critical point simulated by Monte Carlo method
(MCS means Monte Carlo steps). The upper and lower dashed lines
indicate the theoretical values expected from the known and from our
critical exponents, respectively.}
\label{fig1}
\end{figure}

 The MC simulations of Ref.~\cite{IS} allows to determine the fractal
dimensionality $D$ (the largest cluster in the relevant configuration
has the volume $L^D$ where $L$ denotes the linear size of the system)
which is related to the critical exponents by
$\gamma=\nu (2D-d)$ or, which is the same, $\eta=2-\gamma/\nu=d+2-2D$.
In our opinion, this method is better than other more convenient simulation
methods, since it provides the value of $\eta$ as a result of
direct simulation, i.~e., there are no fitting parameters. Besides,
the result is relatively insensitive to the precise value of the critical
coupling (temperature). In Fig.~\ref{fig1} we have shown the average
values of $D$ (the averaging is is made over the MC steps from 1 to 10
(except the initial point), from 11 to 20, and so on) calculated from the
MC data of Ref.~\cite{IS} by measuring deviation
from the line $D=2.48$ in Fig.~8 (of Ref.~\cite{IS}).
If properly treated,
these simulation data confirm the value of $\eta$ about $1/8$
(or $D=2.4375$) consistent with our prediction $\gamma=5/4$ and
$\nu=2/3$, as it is evident from Fig.~\ref{fig1}. The value
$D=2.46 \pm 0.01$ reported in Ref.~\cite{IS} seems to be determined
from the upper MC points (Fig.~8 in Ref.~\cite{IS}) only which are
closer to the known theoretical prediction $D=2.48$.

\begin{figure}
\centerline{\psfig{figure=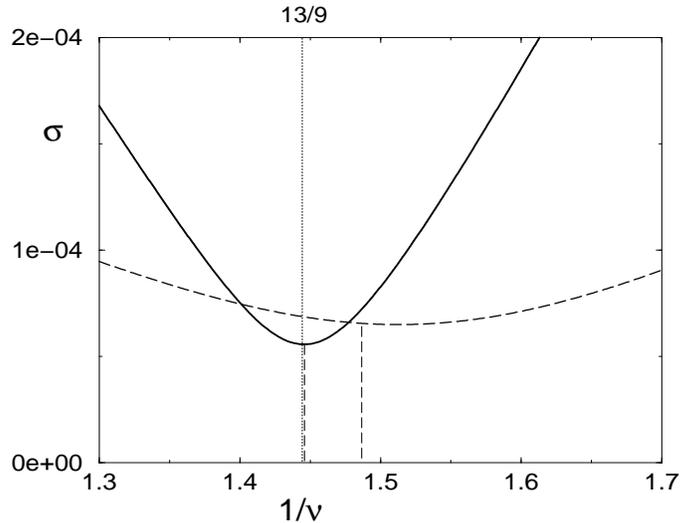,width=9cm,height=7cm,angle=-90}}
\caption{\small The standard deviation $\sigma$
vs the value of $1/\nu$ used in the least--squares fit of
the finite--size scaling curve to the simulated results including
11 data points (solid curve) and 9 data points (dashed curve).
Minimum of the solid curve, shown by a vertical dashed line,
corresponds to the best fit $1/\nu=1.4457$ which is close to
our theoretical value $13/9$ indicated by a vertical dotted line.
Other vertical dashed line indicates the value $1.487$ proposed
by authors of Ref.~\cite{SM}. }
\label{fig2}
\end{figure}

  As regards the MC simulations
of the energy density $E$ of $XY$ model~\cite{SM} at the critical point,
the true picture can be reconstructed from the simulated values listed
in Tab.~I of Ref.~\cite{SM}. Since all the values of $E$ are of comparable
 accuracy, it is purposeful to use the least--squares method to find the
optimum value of $1/\nu$ by fitting the MC data to the prediction of the
finite--size scaling theory
\begin{equation} \label{eq:anal}
E(L)=E_0+E_1 L^{{1 \over \nu}-d} \;,
\end{equation}
where $E(L)$ is the energy density at the critical temperature
$T_{\lambda}$ depending on the linear size of the system $L$.
The standard deviation of the simulated data points from the
analytical curve~(\ref{eq:anal}) can be easily calculated for
any given value of $1/\nu$ with the parameters $E_0$ and
$E_1$ corresponding to the least--squares fit. The result is shown
in Fig.~\ref{fig2}.
The thick solid curve is calculated including all 11 data points
($L$=10, 15, 20, 25, 30, 35, 40, 45, 50, 60, 80), whereas
the dashed line -- including 9 data points (except $L$=10, 15)
used for the fitting in Ref.~\cite{SM}.
Minimum of the solid curve, shown by a vertical dashed line,
corresponds to the best fit $1/\nu=1.4457$ which comes very close to
our theoretical value $13/9$ ( provided by~(\ref{eq:result}) at
$j=1$ and $m=3$) indicated by a vertical dotted line. We have estimated
the statistical error of this MC result about $\pm 0.007$ by
comparing the best fits for several random data sets. Different data
sets have been generated from the original one by omitting some
data points with $10<L<80$. We have found it unreasonable
to omit the data points with two smaller sizes, as it has been
proposed in Ref.~\cite{SM}, since the result in this case becomes
very poorly defined, i.~e., the dashed curve in Fig.~\ref{fig2} has a
very broad minimum. Besides, there is no reason to omit the smallest
sizes, since the analytical curve~(\ref{eq:anal}) excellently fit
all the data points and the standard deviation for 11 data points
is even smaller than that for 9 data points (see Fig.~\ref{fig2}).
The possible systematical error
due to the inaccuracy in the critical temperature
$T_{\lambda}=2.2017 \pm 0.0005$ (the error bars are taken
from the source of this estimation~\cite{Janke}) used in the
simulations~\cite{SM} has been evaluated $\pm 0.017$ by comparing
the simulation results at $T_{\lambda}$ values  2.2012, 2.2017,
and 2.2022. In this case the values of the energy density at
a slightly shifted temperature have been calculated from the
specific heat data given in Tab.~I of Ref.~\cite{SM}.
In such a way, our final estimate from the original MC data
of Ref.~\cite{SM} is $1/\nu=1.446 \pm 0.025$ in a good agreement
with our theoretical value $13/9=1.444...$ and in a clear disagreement
with the usual (RG) prediction about $1.492$. One can only wonder
where the value $1.487$ proposed in Ref.~\cite{SM} comes from.
It does not correspond neither to the best fit for 11 data points
nor to that for 9 data points, as it is evident from
Fig.~\ref{fig2}. The values of $1/\nu$ and $\alpha/\nu$
cannot be determined independently from the discussed here
energy density data.
One of them have to be calculated from the scaling relation
$\alpha/\nu + d =2/\nu$. If authors of Ref.~\cite{SM} were able to
determine $1/\nu$ with $\pm 0.081$ accuracy, then they should be
able to find $\alpha/\nu$ with $\pm 0.162$ accuracy. In this aspect,
the estimate $\alpha/\nu = -0.0258 \pm 0.0075$ given by the
authors looks more than strange.

\begin{figure}
\centerline{\psfig{figure=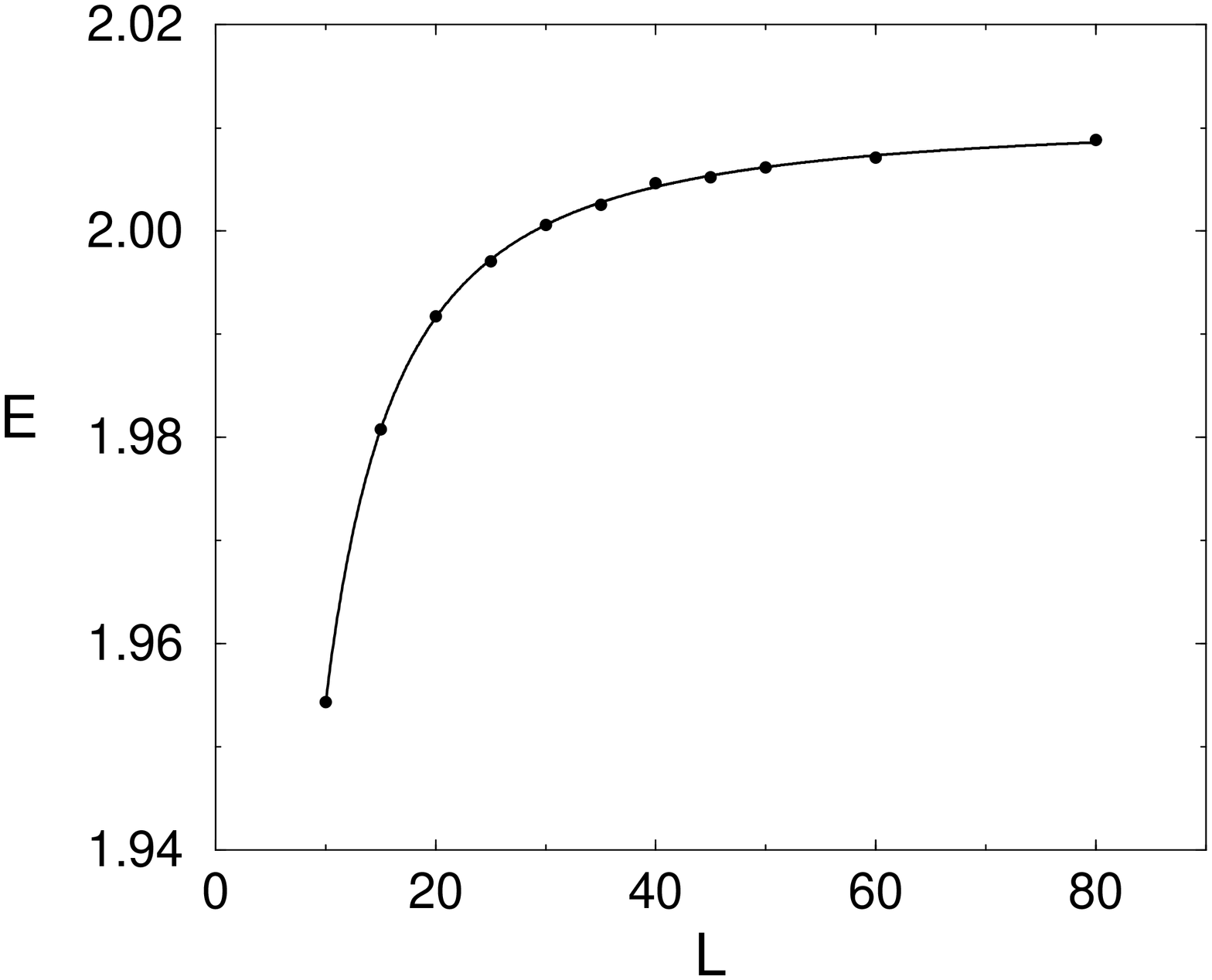,width=6.5cm,height=5.5cm,angle=-90}
\hspace*{1ex} \psfig{figure=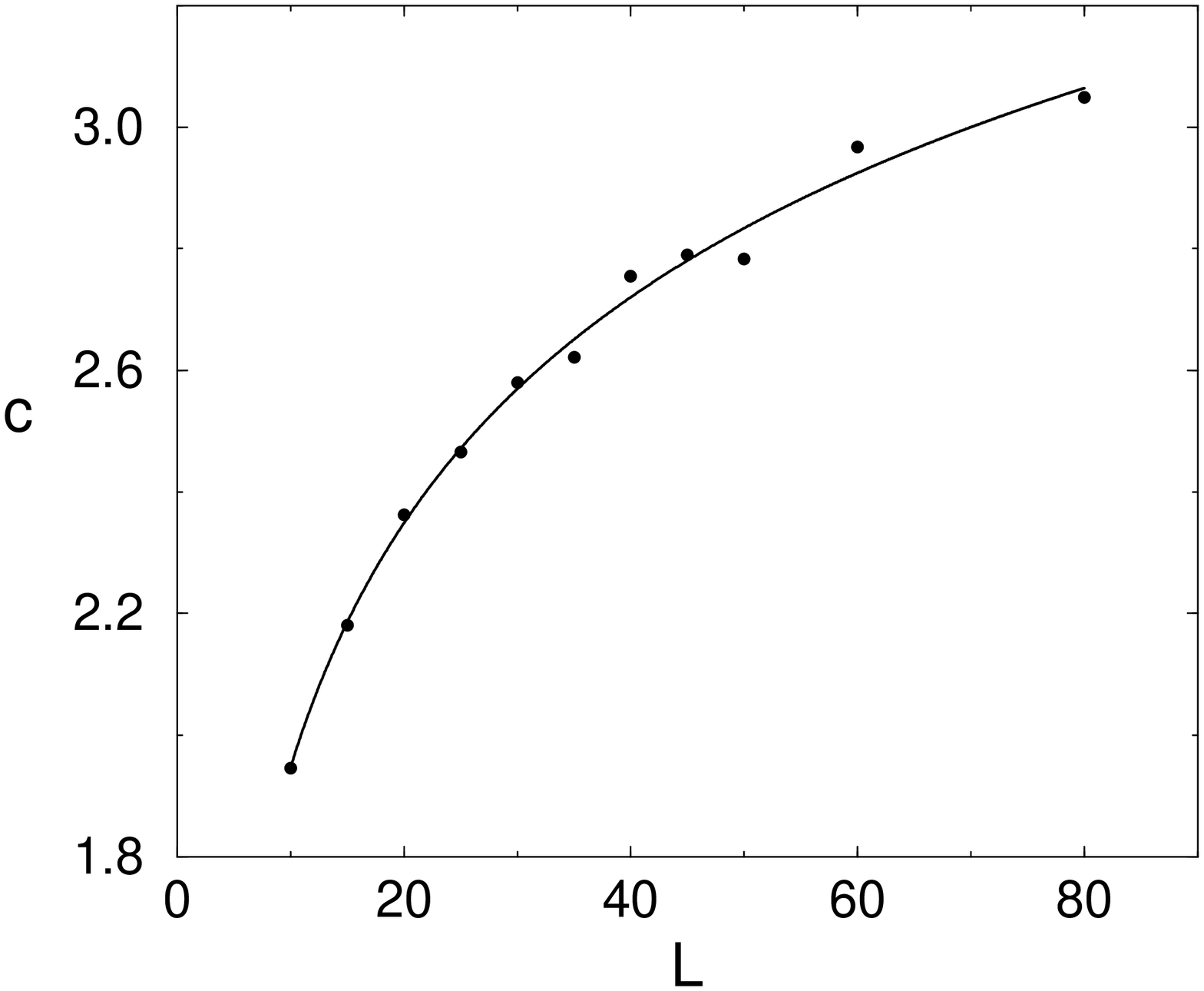,width=6.5cm,height=5.5cm,angle=-90} }
\caption{\small Our fits to the original MC data of Ref.~\cite{SM}
for the energy density (left) and for the specific heat (right)
depending on the linear size of the system $L$. }
\label{fig21}
\end{figure}

In Fig.~\ref{fig21} we have shown our fits to the MC data for the energy
density $E(L)=2.0108-2.0286 \, L^{-14/9}$
and for the specific heat $c(L)=7.360-6.990 \, L^{-1/9}$.
They do not look worse than those in Ref.~\cite{SM},
 but our fit for $c(L)$ seems to be better.

\begin{figure}
\centerline{\psfig{figure=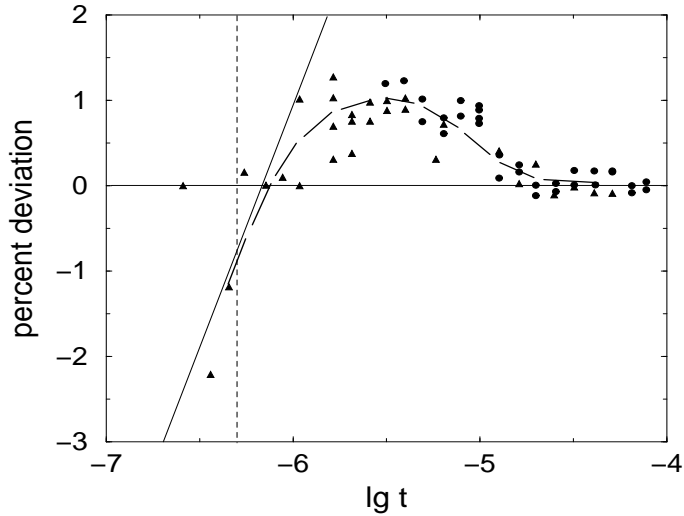,width=9cm,height=7cm,angle=-90}}
\caption{\small Percent deviation of the experimental $\rho_s /\rho$
data~\cite{GA} from the expected theoretical relation (Eq.~(\ref{eq:fit})
at $\delta=0$). The stright line shows the slope of this plot at
the value of $t$, equal to $5 \cdot 10^{-7}$, indicated by a vertical
dashed line. }
\label{fig3}
\end{figure}

  One believes that the value of critical exponent $\nu$
about $0.67$, predicted by the RG theory at $n=2$, is well
confirmed by very accurate measurements of the superfluid fraction
$\rho_s/\rho = y$ in $^4 He$. This is not true, since in reality
these experiments~\cite{GA}
provide a good evidence that the effective critical exponent
$\nu_{eff}(t)=\partial (\ln y)/ \partial (\ln t)$
remarkably increases when the reduced temperature
$t=(T_{\lambda}-T)/T_{\lambda}$ (where $T_{\lambda}$ is the
critical temperature) is decreased below $10^{-5}$.
According to Ref.~\cite{GA}, $\rho_s/\rho$ is given by
\begin{equation} \label{eq:fit}
\rho_s/\rho= y(t) =k_0(1+k_1 t) (1+D_{\rho} t^{\Delta}) t^{\zeta}
\times (1+\delta(t)) \; ,
\end{equation}
where $k_0$, $k_1$, $D_{\rho}$, and $\zeta$ are the fitting
parameters, $\Delta=0.5$ is supposed to be the correction--to scaling
exponent, and $\delta(t)$ is the measured relative deviation
from the expected theoretical expression obtained by setting
$\delta(t)=0$. The percent deviation discussed in Ref.~\cite{GA}
is 100 times $\delta(t)$. From Eq.~(\ref{eq:fit}) we obtain
\begin{equation} \label{eq:ffit}
\nu_{eff}(t)= \zeta + \frac{k_1 t}{1+k_1 t}
+ \frac{\Delta D_{\rho} t^{\Delta}} {1+D_{\rho} t^{\Delta}}
+ \frac{1}{1+ \delta(t)}
\times \frac{\partial \delta(t)}{\partial (\ln t)} \; .
\end{equation}
For the values of $t$ as small as $t<10^{-5}$ and for $\delta(t) \ll 1$
 Eq.~(\ref{eq:ffit}) with the fitting parameters $\zeta=0.6705$, $k_0=2.38$,
$k_1=-1.74$, and $D_{\rho}=0.396$ used in Ref.~\cite{GA} reduces to
\begin{equation} \label{eq:fi}
\nu_{eff}(t) \simeq \zeta + \partial \delta(t) / \partial (\ln t) \;.
\end{equation}
The second term in this equation is proportional to the slope of the
percent deviation plot $100 \, \delta(t)$ vs $\ln t$ or $\lg t$
(the decimal logarithm) in Figs.~2 and 3 of Ref.~\cite{GA}. We have read the
experimental data from Fig.~2 in Ref.~\cite{GA} within the region
$t<10^{-4}$ and have depicted them in Fig.~\ref{fig3}.
Almost all the data points
with a reasonable accuracy fit the smooth curve $\delta(t)$ vs $\lg t$
(dashed line) having a maximum at about $\lg t= -5.5$. It means that
$\partial^2  \delta(t) / \partial (\ln t)^2$ is negative within some
region around the maximum, i.~e., according to~(\ref{eq:fi})
the effective critical exponent $\nu_{eff}(t)$ increases if $t$ is decreased.
We have roughly estimated and have shown by  stright line the
slope of this curve at $t=t^* =5 \cdot 10^{-7}$ ($t^*$ value is indicated
in Fig.~\ref{fig3} by vertical dashed line). From this we obtain
$\partial \delta(t)/\partial(\ln t) \approx 0.025$.
This result depends on
the shift in the experimentally determined $T_{\lambda}$ value.
To obtain a more reliable estimate, we have performed the same manipulations
with the data depicted in Fig.~3 of Ref.~\cite{GA} corresponding to
$T_{\lambda}$ shifted by $\pm 20 nK$, and have obtained the values of
$\partial \delta(t)/\partial(\ln t)$ about $0.03$ and $0.015$, respectively.
Our final result $0.0233 \pm 0.0083$ for this derivative at $t=t^*$ has
been obtained by averaging over the three above discussed
estimates ($0.015$, $0.025$, and $0.03$) with the error bars large enough
to include all these values. According to this, from
Eq.~(\ref{eq:fi}) with $\zeta =0.6705$ we obtain
$\nu_{eff}(t^*)=0.694 \pm 0.009$ which, again, is in a good agreement with
the value $\nu=9/13 \simeq 0.6923$ provided by Eq.~(\ref{eq:result})
at $j=1$ and $m=3$ and in a disagreement with the RG predictions.

\section{Conclusions}

We have proposed a novel method in critical phenomena
(Sect.~\ref{sec:my},~\ref{sec:myexp}) which is based on the
grouping of Feynman diagrams in $\varphi^4$ model with $O(n)$
symmetry. As a result,
equations for calculation of the two--point correlation
function have been obtained containing an infinite, but converging
perturbation sum. It has been shown that these equations
allow, in principle, to find the exact critical exponents.
In distinction to the usual renormalization group approach,
our predictions are based not on evaluation of some of the
first terms in the perturbation expansion, but on the proof
of relevant scaling properties for the whole
sum, which is possible due to the actually proposed
reorganization of the perturbation theory.

 Based on this scaling analysis, we have derived a set of
possible values for the exact critical
exponents~(\ref{eq:result}).
A disagreement with the actually accepted values of the
critical exponents in three dimensions has been revealed.
However, we argue that our result is correct since,
in distinction to the usual treatment critised in
Sec.~\ref{sec:RG} and~\ref{sec:random},
our method is faultless from the mathematical point
of view. Some assumptions have been made, but they look innocent
and have been well motivated. Besides, our method,
being equally valid in two and three dimensions,
reproduces the known exact critical exponents at $d=2$.

 A comparison of results has been made in
Sect.~\ref{sec:compare}, showing that in some
cases, at least, our predictions are in accurate agreement with
properly treated MC simulation data as well as with experiments.
More comparison with MC data is in progress.

 In summary, we conclude the following.
\begin{itemize}

\item[1.] The conventional method of the perturbative RG theory
is contradictory and, therefore, cannot give correct values of
critical exponents.
\item[2.] Our equations, derived in Sec.~\ref{sec:my} by
grouping of the Feynman diagrams, allow to find the exact critical
exponents for $\varphi^4$ model with $O(n)$ symmetry ($n \ge 1$)
by proving all the relevant scaling properties of the asymptotic
solutions at $T=T_c$ and $T \to T_c$, not cutting the perturbation
series. These scaling properties have been proven in
Sec.~\ref{sec:myexp}.
\item[3.] In the cases of the
second--order phase transition with spontaneous long--range
ordering, all possible values for the exact critical exponents
at $d=2, 3$ and $n=1, 2, 3$, etc. are given by Eq.~(\ref{eq:result}).
\item[4.] At $m=3$ and $j=0$ our result~(\ref{eq:result}) reproduces
the known exact critical exponents in two dimensions at $n=1$
($\gamma=7/4$ and $\nu=1$).
Based on the idea that $d$ may be considered as a continuous
parameter, we conclude that $\gamma=5/4$ and $\nu=2/3$
are the true (exact) values at $d=3$ and $n=1$.
\item[5.] The comparison with Monte--Carlo data in
Sec.~\ref{sec:compare} well confirms the hypothesis that
$\gamma=17/13$ and $\nu=9/13$ (corresponding to $m=3$, $j=1$)
are the true values of the critical exponents at $d=3$
and $n=2$.
\end{itemize}

\textbf{\Large Appendix A. \hspace{1ex} Large--order estimation }
\vspace*{1ex}

   Let us consider coupled diagrams, constructed of vertices
{\mbox \wwv,} which are skeleton diagrams with respect to both solid and
waved lines (i.~e., do not contain parts \selfe and/or {\mbox
\wblock).} At large $m$ most of the coupled skeleton diagrams of
the $(m+1)$--th order can be constructed by coupling vertex \wwv to a diagram
\onebl  of the $m$--th order with four outer lines obtained by breaking two of
$2m$ solid lines in a coupled skeleton
diagram of the $m$--th order. Since at large $m$ the lines in most
cases of coupled diagrams are arranged randomly (i.~e., the symmetry
elements are not important) this yields approximately $6 m^2$
(relevant to $m \cdot (2m-1)$ possible ways to break the solid lines, and
$3$ different ways to couple the vertex to get a skeleton diagram)
topologically nonequivalent diagrams of the $(m+1)$--th
order corresponding to a given diagram of the $m$--th order. However, different diagrams of the
$m$--th order yield the same diagram of the $(m+1)$--th order. Really, we can
subtract any of the $m+1$ vertices \wwv from a (random) coupled diagram of
the $(m+1)$--th order and obtain a diagram \onebl of
the $m$--th order. $m+1$ reverse procedures represent $m+1$ different ways of
how to obtain this coupled diagram of the $(m+1)$--th order from diagrams of
the $m$--th order. Thus, the total number of the above defined coupled skeleton
diagrams of the $(m+1)$--th order
exceeds the number of corresponding diagrams of the $m$--th order approximately
$6 m$ times, or the number of diagrams at large  $m$  increases approximately
as $6^m m!$.

   Let us consider other possibilities to construct coupled skeleton diagrams
of the \linebreak $(m+1)$--th order. We can couple a vertex \wwv to a diagram
obtained by
breaking two solid lines in a non skeleton diagram. The simplest of such
diagrams \twobl and \twowbl are constructed of two above defined skeleton
diagrams of orders $m_1$  and $m_2$ with $m_1+m_2=m$. The number of such
diagrams relative to the number of the diagrams \onebl obtained from the
skeleton diagrams of the $m$--th order tends to zero at $m \to \infty$ as
$\left( m_1^2 \, m_2^2 \, m_1! \, m_2! \right) / \left( m^2 m! \right)$. More
complicated diagrams like \chabl (where any of the neighbouring blocks can be
connected by waved lines instead of the solid lines)  also can be included.
However, this gives only a second--order correction. Other
kind of diagrams among those obtained by breaking two solid lines in a non
skeleton diagram, e.~g., {\mbox \thrblo,} are not included because the coupling
of vertex in this case does not yield a skeleton diagram. In such a way, the
number of possible constructions is increased  insignificantly if non skeleton
diagrams of the $m$--th order are included. So, factor $6 m$ is corrected
only slightly. For a more accurate estimation this factor is
replaced by $6m +o(1)$, which allows to conclude that the
number of diagrams of the $m$--th order in Eq.~(\ref{eq:ee})
increases with $m$ like
$const \cdot m^{\gamma} \, 6^m m!$, where $\gamma$ is a constant.
Note that the diagrams of~(\ref{eq:ee}) are obtained from
the actually discussed coupled skeleton diagrams by breaking a
waved line, replacing this line by a (broken) dashed line.

Consider now a random high--order diagram $\Sigma_{m,i}({\bf
q},\zeta)$ of Eq.~(\ref{eq:ee}), i.~e., the $i$--th diagram among
those of the $m$--th order. Calculation of the sum over wave vectors
and order--parameter--components for such a diagram
is analogous to finding of the statistical sum of a random
lattice formed by the lines of this diagram. Two conditions must
be fulfilled: first, the sum of the wave vectors coming into any of
the kinks is zero, the wave vectors of the outer lines
being fixed (${\bf q}$ and $-{\bf q}$), and, second, the same
order--parameter--component is related to solid lines of a closed
loop. These conditions represent certain interaction between
the lines of the diagram.
The "interaction energy" is zero if they are satisfied,
and $\infty$ otherwise.
Besides, each line has his ownw "energy". If normalized to $T$,
this energy is equal to minus logarithm
of the absolute value of the factor related to the corresponding line
in the diagram notation. This analogy can be used in calculation of
$\mid \Sigma_{m,i}({\bf q},\zeta) \mid$.
In a typical case of large $m$ this yields
$\Sigma_{m,i}({\bf q},\zeta) \sim -(-1)^m b^m$, where $b^m$ with
$b>0$ represents the absolute value and $-(-1)^m$ -- the sign of this term.
This result remains true (with new value of $b$) if the combinatorial
factor is included, since the latter is $4 \cdot 8^m$ for any random diagram
having no symmetry elements (see Sect.~\ref{sec:notation}).
\vspace*{1ex}

\textbf{\Large Appendix B. \hspace{1ex} Non--perturbative analysis }
\vspace*{1ex}

For convenience, here we consider the one--component case $n=1$,
since the specific (natural) value of $n$ is irrelevant in the present scaling
analysis. The Hamiltonian of the considered model
$$
H/T = \int \left[ r_0 \varphi^2 ({\bf x})
+c {\left( \nabla \varphi({\bf x}) \right)}^2
+ u \varphi^4 ({\bf x}) \right] d {\bf x} \eqno(B1)
$$
in the Fourier representation is
$$
H/T = \sum\limits_{\bf k} \left( r_0+c {\bf k}^2 \right)
{\mid \varphi_{\bf k} \mid}^2 + u V^{-1}
\sum\limits_{{\bf k}_1,{\bf k}_2,{\bf k}_3}
\varphi_{{\bf k}_1} \varphi_{{\bf k}_2} \varphi_{{\bf k}_3}
\varphi_{-{\bf k}_1-{\bf k}_2-{\bf k}_3} \; . \eqno(B2)
$$
Summation in Eq.~(B2) is performed within
$k<k_0$ (here $k_0 \equiv \Lambda$). Points in the ${\bf k}$--space are
separated by a distance
$2 \pi / L$, where $L$ is the linear size of the system with volume
$V=L^d$ at $d<4$. By a formal change of variables
$\Psi_{\bf p}=u^{\alpha} c^{-d \alpha/2} \varphi_{\bf k}$,
${\bf p}=c^{2 \alpha} u^{- \alpha} {\bf k}$,
$p_0=c^{2 \alpha} u^{- \alpha} k_0$, and
$R=r_0 \, c^{d \alpha} u^{-2 \alpha}$, where $\alpha=1/(4-d)$
we obtain
$$
H/T = \sum\limits_{\bf p} \left( R+{\bf p}^2 \right)
{\mid \Psi_{\bf p} \mid}^2 + V_{1}^{-1}
\sum\limits_{{\bf p}_1,{\bf p}_2,{\bf p}_3}
\Psi_{{\bf p}_1} \Psi_{{\bf p}_2} \Psi_{{\bf p}_3}
\Psi_{-{\bf p}_1-{\bf p}_2-{\bf p}_3} \eqno(B3)
$$
where $V_1=L_1^d$  with $L_1=L \cdot c^{-2 \alpha} u^{\alpha}$.
In Eq.~(B3) the summation is performed within $p<p_0$. Points are
separated by a distance $2 \pi / L_1$ . According to equations~(B2)
and (B3), we have (fluctuation modes with $k>k_0$ are excluded)
$$
G({\bf k})= \left< {\mid \varphi_{\bf k} \mid}^2 \right>
=c^{d \alpha} u^{-2 \alpha} \left< {\mid \Psi_{\bf p} \mid}^2 \right>
=c^{d \alpha} u^{-2 \alpha} g({\bf p}, p_0, R) \eqno(B4)
$$
where, at a fixed  $d$  and $V_1 \to \infty$, $g$ is a function of the given
arguments only. The latter is true since the only parameters contained in
Hamiltonian~(B3) are $p_0$, $R$, and $V_1$. For the same reason, the critical
value of $R$ depends merely on $p_0$, and the correlation function at the
critical point is
$$
G^*({\bf k})=c^{d \alpha} u^{-2 \alpha} g^*({\bf p}, p_0) \; .
\eqno(B5)
$$
It follows from Eq.~(B5) that the width of the critical region for $p$ is
$p_{crit}= \phi(p_0)$ where $\phi(z)$ is a single--argument function.
Here both $d$ and the accuracy parameter $\varepsilon$ in
Eq.~(\ref{eq:kcr}) are considered as fixed quantities.
According to the relation between
${\bf p}$ and ${\bf k}$, we have $p_{crit}=c^{2 \alpha} u^{-\alpha} k_{crit}$,
which yields
$$
k_{crit}= c^{-2 \alpha} u^{\alpha}
\phi \left( c^{2 \alpha} u^{-\alpha} k_0 \right) \; . \eqno(B6)
$$
Similarly, by using Eq.~(B5) we obtain an exact scaling relation
for the asymptotic expansion at $k \to 0$, i.~e.,
$$
G^*({\bf k})= \sum\limits_l b_l \, k^{-\lambda_l} \; , \eqno(B7)
$$
where $\lambda_0 \equiv \lambda=2-\eta$ and terms with $l>0$
are corrections to scaling with the amplitudes represented
in the scaled form
$$
b_l = B_l \left( c^{2 \alpha} u^{-\alpha} k_0 \right)
\cdot c^{\alpha (d-2 \lambda_l)} u^{\alpha (\lambda_l-2)} \; .
\eqno(B8)
$$
It is evident from Eq.~(B6) that $k_{crit}$ can exponentially (like
$\exp \left( -u^{-\sigma} \right)$ with $\sigma >0$) tend to zero at
$u \to 0$ only if the function $\phi(z)$ decreases exponentially at
$z \to \infty$. Let us now consider the
behavior of $k_{crit}$ at $k_0 \to \infty$ at fixed $c$ and $u$. We conclude
immediately: if $k_{crit}$ decreases exponentially at $u \to 0$, then this
quantity behaves in the same way at $k_0 \to \infty$. Physically, this
means that the width of the critical region is strongly affected by
short--wave fluctuations and the effect dramatically
(exponentially) increases with shortening of the wavelength.
The latter is rather unphysical,
since the critical behavior is well known to be governed by long--wave
fluctuations. Thus, $k_{crit}$ cannot decrease exponentially (or,
in general, faster than $u^s$ at any $s>0$) at
$u \to 0$, which means that
$\lim\limits_{u \to 0} \left( u^s/k_{crit}(u) \right) =0$
holds at large enough $s$. Based on Eq.~(B8), the same conclusion
can be made for the amplitudes $b_l$.


\begin{thebibliography}{}

\bibitem{Baxter} Rodney J.~Baxter, Exactly Solved Models in
Statistical Mechanics, Academic Press, London, 1989
\bibitem{WF} K.G.~Wilson, M.E.~Fisher,
Phys.Rev.Lett. {\bf 28} (1972) 240
\bibitem{Ma} Shang--Keng Ma, Modern Theory of Critical
Phenomena, W.A.~Benjamin, Inc., New York, 1976
\bibitem{Justin} J.~Zinn--Justin, Quantum Field Theory and
Critical Phenomena, Clarendon Press, Oxford, 1996
\bibitem{GE} A.J.~Guttman, I.G.~Enting, J.~Phys.~A {\bf 27}
(1994) 8007
\bibitem{BC} P.~Butera, M.~Comi, Phys.~Rev. B {\bf 56} (1997) 8212
\bibitem{Wilson} K.G.~Wilson, Phys.~Rev.~Lett. {\bf 28}
(1972) 548
\bibitem{BGZN} E.~Br\'ezin, J.C.~Le~Guillou, J.~Zinn--Justin,
B.G.~Nickel, Phys.~Lett. A {\bf 44} (1973) 227
\bibitem{BWW} E.~Br\'ezin, D.J.~Wallace, K.G.~Wilson,
Phys.~Rev.~B {\bf 7} (1973) 232
\bibitem{VTK} A.A.~Vladimirov, D.I.~Kazakov, O.V.~Tarasov,
Sov.~Phys.~JETP {\bf 50} (1979) 521
\bibitem{CKT} K.G.~Chetyrkin, A.L.~Kataev, F.V.~Tkachov,
Phys.~Lett.~B {\bf 99} (1981) 147
\bibitem{GLT} S.G.~Gorishny, S.A.~Larin, F.V.~Tkachov,
Phys.~Lett.~A {\bf 101} (1984) 120
\bibitem{KNSCL} H.~Kleinert, J.~Neu, V.~Schulte--Frohlinde, K.G.~Chetyrkin,
S.A. Larin, Phys.~Lett.~B {\bf 272} (1991) 39, Erratum {\bf 319} (1993) 545
\bibitem{ILF} Yu.M.~Ivanchenko, A.A.~Lisyanskii, A.E.~Filipov,
Phys.~Lett.~A {\bf 150} (1990) 100
\bibitem{Parisi} G.~Parisi, Carg\`{e}se Lectures 1973,
published in J.~Stat.~Phys. {\bf 23} (1980) 49
\bibitem{BNGM} G.A.~Baker, B.G.~Nickel, M.S.~Green,
D.I.~Meiron, Phys.~Rev.~Lett {\bf 36} (1976) 1351
\bibitem{BNM} G.A.~Baker, B.G.~Nickel, D.I.~Meiron,
Phys.~Rev.~B {\bf 17} (1978) 1365
\bibitem{GJ77} J.C.~Le~Guillou, J.~Zinn--Justin,
Phys.~Rev.~Lett. {\bf 39} (1977) 95
\bibitem{GJ80} J.C.~Le~Guillou, J.~Zinn--Justin,
Phys.~Rev.~B {\bf 21} (1980) 3976
\bibitem{GuiJ} R.~Guida, J.~Zinn--Justin, J.~Phys.~A {\bf 31}
(1998) 8103
\bibitem{AS} S.A.~Antonenko, A.I.~Sokolov,
Phys.~Rev.~E {\bf 51} (1995) 1894
\bibitem{EMS} J.P.~Eckmann, J.~Magnen, R.~S\`en\`eor,
Commun.~Math.~Phys. {\bf 39} (1975) 251
\bibitem{FO} J.S.~Feldman, K.~Osterwald, Ann.~Phys.~(NY)
{\bf 97} (1976) 80
\bibitem{EE} J.P.~Eckman, H.~Epstein, Commun.~Math.~Phys.
{\bf 68} (1979) 245
\bibitem{WJ} H.~Weigel, W.~Janke, Annalen der Physik,
{\bf 7} (1998) 575
\bibitem{Tseskis} A.L.~Tseskis, Zh.~Eksp.~Teor.~Fiz.
{\bf 102} (1992) 508
\bibitem{MK1} I.~Madzhulis, J.~Kaupu\v{z}s,
Phys.~Stat.~Sol. (b) {\bf 175} (1993) 307
\bibitem{MK2} I.~Madzhulis, J.~Kaupu\v{z}s,
Phys.~Stat.~Sol. (b) {\bf 176} (1993) 41
\bibitem{IS} N.~Ito, M.~Suzuki, Progress of Theoretical
Physics {\bf 77} (1987) 1391
\bibitem{SM} N.~Schultka, E.~Manousakis, Phys.~Rev.~B
{\bf 52} (1995) 7258
\bibitem{Janke} W.~Janke, Phys.~Lett.~A {\bf 148} (1990) 306
\bibitem{GA} L.S.~Goldner, G.~Ahlers, Phys.~Rev.~B
{\bf 45} (1992) 13129
\end{thebibliography}
\end{document}